\documentclass[twocolumn,aps,prx,superscriptaddress]{revtex4-1}
\usepackage[utf8]{inputenc}
\usepackage{color}
\usepackage{verbatim}
\usepackage{amsmath}
\usepackage{amssymb}
\usepackage{graphicx}
\usepackage[unicode=true,
 bookmarks=false,
 breaklinks=false,pdfborder={0 0 1},backref=false,colorlinks=true]
 {hyperref}
\hypersetup{
 linkcolor=blue,citecolor=blue,urlcolor=blue,linkcolor=blue,citecolor=blue,urlcolor=blue}

\makeatletter
\usepackage{color}
\usepackage{verbatim}

\usepackage{braket}

\newcommand{\tr}{\mathop{\mathrm{tr}}}
\newcommand{\abs}[1]{\left\lvert #1 \right\rvert}

\makeatother

\begin{document}

\title{Quantum Non-demolition Measurement of a Many-Body Hamiltonian}

\author{Dayou Yang}

\author{Andrey Grankin}

\author{Lukas M. Sieberer}

\author{Denis V. Vasilyev}

\author{Peter Zoller}

\email{peter.zoller@uibk.ac.at}

\affiliation{Center for Quantum Physics, University of Innsbruck, 6020 Innsbruck,
Austria}

\affiliation{Institute for Quantum Optics and Quantum Information of the Austrian
Academy of Sciences, 6020 Innsbruck, Austria}

\date{\today}
\begin{abstract}
In an ideal quantum measurement, the wave function of a quantum system collapses to an eigenstate of the measured observable, and the corresponding eigenvalue determines the measurement outcome. If the observable commutes with the system Hamiltonian, repeated measurements yield the same result and thus minimally disturb the system. Seminal quantum optics experiments have achieved such quantum non-demolition (QND) measurements of systems with few degrees of freedom. In contrast, here we describe how the QND measurement of a complex many-body observable, the Hamiltonian of an interacting many-body system, can be implemented in a trapped-ion analog quantum simulator. Through a single-shot measurement, the many-body system is prepared in a narrow band of (highly excited) energy eigenstates, and potentially even a single eigenstate. Our QND scheme, which can be carried over to other platforms of quantum simulation, provides a framework to investigate experimentally fundamental aspects of equilibrium and non-equilibrium statistical physics including the eigenstate thermalization hypothesis and quantum fluctuation relations.
  \end{abstract}
\maketitle


\section{Introduction}
Recent experimental advances provide intriguing opportunities in the
preparation, manipulation, and measurement of the quantum state of engineered
complex many-body systems. This includes the ability to address individual sites
of lattice systems enabling single-shot read-out of single-particle observables,
as demonstrated by the quantum gas microscope for atoms in optical
lattices~\cite{Parsons1253,Boll1257}, single-spin or qubit read-out of trapped
ions~\cite{Brydges2018,Garttner2017,Negnevitsky:2018aa,Rossnagel325,Landsman:2019aa}
and Rydberg tweezers
arrays~\cite{Norcia2018,Cooper2018,Keesling:2019aa,Barredo:2018aa,PhysRevLett.122.143002},
and superconducting qubits~\cite{Barends2016,PhysRevX.8.021003}.  In contrast,
we are interested below in developing single-shot measurements of many-body
observables such as the Hamiltonian $\hat{H}$ of an interacting many-body
system. For an isolated quantum system, $\hat{H}$ represents a QND observable,
and our goal is to implement a QND measurement of energy of a quantum many-body
system in an analog simulator setting. We note that quantum optics provides with
several examples of QND measurements; however these have so far been confined to
observables representing few quantum degrees of
freedom~\cite{Gleyzes2007,Johnson2010,Volz2011,PhysRevX.8.021003,Hacohen-Gourgy:2016aa,PhysRevLett.99.120502,Eckert:2007aa}.

Developing QND measurement of a many-body Hamiltonian $\hat{H}$ provides us
first of all with the unique opportunity to distill---in a single run of the
experiment---an energy eigenstate $\ket{\ell}$ from an initial, possibly mixed,
or finite temperature state, by observing in particular run the energy
eigenvalue $E_{\ell}$. In case of measurement with finite resolution, this will
prepare states in a narrow energy window, reminiscent of a microcanonical
ensemble. We emphasize that state preparation by measurement is intrinsically
probabilistic, i.e., will vary from shot to shot, reflecting the population
distribution.  Furthermore, this provides us with a tool to determine
populations and population distributions of (excited) energy eigenstates, as
required in, e.g., many-body spectroscopy~\cite{Senko430}. The ability to
prepare and measure (single) energy eigenstates provides us with a unique tool
to address experimentally fundamental problems in quantum statistical physics,
such as the eigenstate thermalization hypothesis
(ETH)~\cite{Deutsch1991,Srednicki1994,Rigol:2008aa}, which asserts that single
energy eigenstates of an (isolated) ergodic system encode thermodynamic
equilibrium properties. Developing the capability to turn QND measurements on
and off allows one to alternate between time periods of free evolution of the
unobserved many-body quantum system, and energy measurement. This allows quantum
feedback in a many-body system conditional to measurement outcomes, and in
particular provides a framework to monitor non-equilibrium dynamics and
processes in quantum thermodynamics~\cite{Campisi2011}, including measurement of
work functions and quantum fluctuation relations
(QFRs)~\cite{PhysRevLett.101.070403,An:2014aa,Cerisola:2017aa}.  These relations
express fundamental constraints on, e.g. the work performed on a quantum system
in an arbitrary non-equilibrium process, imposed by the universal canonical form
of thermal states and the principle of microreversibility.

Our aim below is to develop QND measurement of $\hat{H}$ in physical settings of
analog quantum simulation, in particular exploring the regime of mesoscopic
system sizes. We will demonstrate this in detail for the example of an analog
trapped-ion quantum simulator, realizing a long-range transverse Ising
Hamiltonian and the associated QND measurement. Our implementation in an analog
quantum device should be contrasted to QND measurement of $\hat{H}$ via a phase
estimation algorithm \cite{McArdle2018}, which however requires a universal
(digital) quantum computer.

\begin{figure*}
\includegraphics[width=1\textwidth]{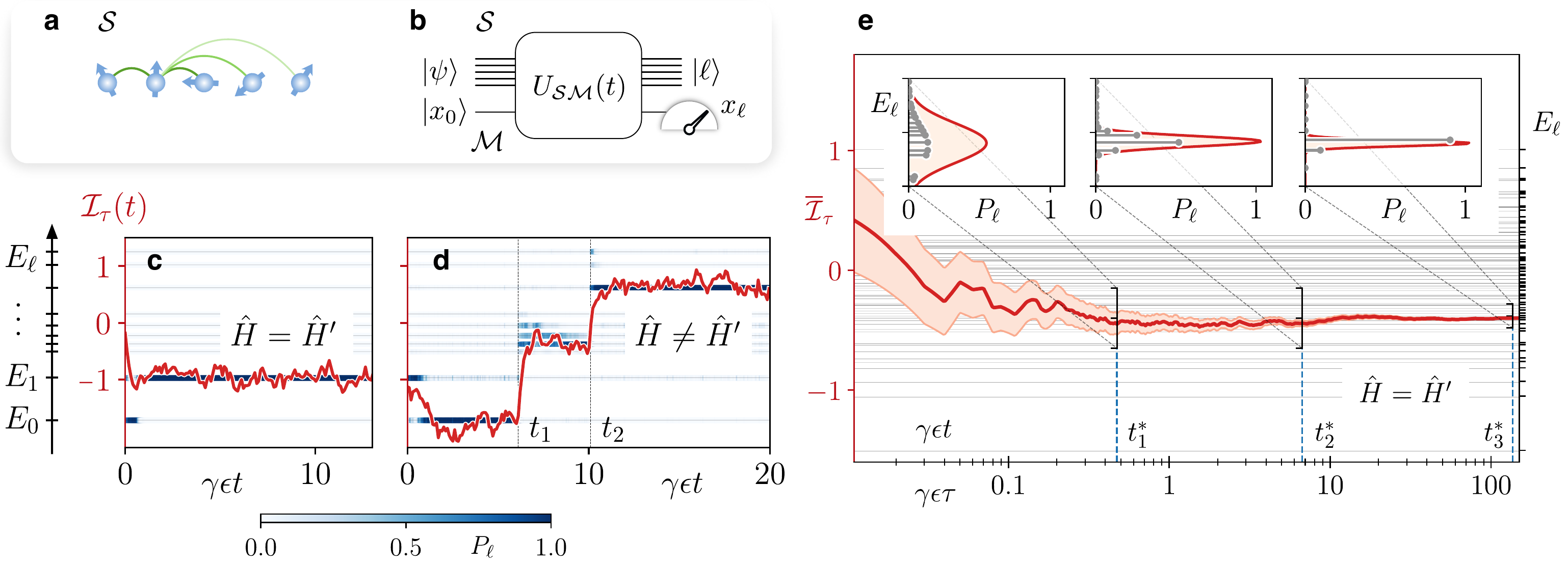}

\caption{QND measurement of a many-body Hamiltonian $\hat{H}$ in a quantum
  simulator setting. The many-body spin system $\mathcal{S}$, shown in (a), is
  entangled with an ancillary system $\mathcal{M}$ (meter) by the unitary
  $\hat{U}_{\mathcal{SM}}(t)=\exp\{-i\int_{0}^{t}dt'[\hat{H}\otimes\mathbb{I}+\vartheta(t')\hat{H}\otimes\hat{P}]\}$.
  (b) Subsequent reading of a meter value
  $x_{\ell}\equiv x_{0}+E_{\ell}\int_{0}^{t}dt'\vartheta(t')$ prepares the
  many-body system in an energy eigenstate $\ket{\ell}$ with the eigenvalue
  $E_{\ell}$. (c) Single trajectory
  simulation~\eqref{eq:SME},~\eqref{eq:current} of an ideal QND measurement for
  the Ising Hamiltonian~\eqref{eq:Ising_H} for $N=5$ spins, $\alpha=1.5$,
  $h/J=1.5$. The window-filtered homodyne current $\mathcal{I}_{\tau}(t)$ (red
  curve) fluctuates around a value corresponding to the eigenenergy prepared by
  the measurement of $\hat{H}$.  The thin horizontal lines show the system
  eigenenergies $E_{\ell}$ and the blue color indicates the conditional
  populations $P_{\ell}(t)$ of the corresponding eigenstates. (d) Observation of
  quantum jumps due to the mismatch of the transverse fields
  $\hat{H}'=\hat{H}+\delta\tilde{h}\sum_{j}\hat{\sigma}_{j}^{z}$ with
  $\protect\smash{\delta\tilde{h}/J=-0.2}$. The filtered photocurrent (red)
  clearly shows sudden jumps between eigenstates at times $t_{1}$ and
  $t_{2}$. (e) Preparation of energy eigenstates or microcanonical ensembles by
  the ideal QND measurement for $N=8$ spins, $\alpha=1.5$, $h/J=0.8$. The
  estimate of the system energy given by the cumulative time-average of the
  homodyne current $\overline{\mathcal{I}}_{\tau}$ (red line) gradually
  converges to a single eigenenergy (grey lines) as averaging time $\tau$
  increases. The corresponding uncertainty (red area) due to shot noise
  decreases as $\sim1/\sqrt{\gamma\epsilon\tau}$.  Inset: conditional population
  of the energy eigenstates (grey points) at times $t_{1}^{*}$, $t_{2}^{*}$, and
  $t_{3}^{*}$ is well captured by gaussian distributions of widths
  $J/\sqrt{4\gamma\epsilon t_{1,2,3}^{*}}$ describing the fluctuations of the
  shot noise averaged over $\tau=t_{1,2,3}^{*}$ (red curve).}
\label{fig:1}
\end{figure*}

\section{Results}
\subsection{QND Measurement of $\hat{H}$}
On a more formal level, we define QND measurement of a many-body Hamiltonian
$\hat{H}$ as an indirect measurement by coupling the system of interest
$\mathcal{S}$, illustrated in Fig.~\ref{fig:1}(a), to an ancillary
system $\mathcal{M}$ as meter. In a first step, the system is entangled
with the meter according to the time evolution $U(t)=\exp(-i\hat{H}_{{\rm QND}}t)$
generated by the QND Hamiltonian 
\begin{equation}
\hat{H}_{{\rm QND}}=\vartheta\hat{H}\otimes\hat{P},\label{eq:H_qnd}
\end{equation}
with coupling strength $\vartheta$ ($\hbar=1$). To be specific and
in light of examples below, we consider here as meter a continuous
variable system with a pair of conjugated quadratures $\hat{X}$ and
$\hat{P}$ obeying the canonical commutation relation $\smash{[\hat{X},\hat{P}]=i}$.
Consider now an initial state of the joint system prepared as $\ket{\Psi}=\ket{\psi}\otimes\ket{x_{0}}$,
where $\ket{\psi}\equiv\sum_{\ell}c_{\ell}\ket\ell$ is a superposition
of energy eigenstates, $\hat{H}\ket\ell=E_{\ell}\ket\ell,$ and $\ket{x_{0}}$
is an (improper) eigenstate of $\hat{X}$ (or squeezed state). We
obtain for the time-evolved state $\ket{\Psi(t)}=\hat{U}(t)\ket{\psi}\otimes\ket{x_{0}}=\sum_{\ell}c_{\ell}\ket\ell\otimes\ket{x_{0}+\vartheta E_{\ell}t}$.
Reading the meter as $x_{\ell}\equiv x_{0}+\vartheta E_{\ell}t$,
and thus measuring the eigenvalue $E_{\ell}$, will prepare the system
in $\ket{\ell}$ (or in the relevant subspace in case of degeneracies).
The probability for obtaining the particular measurement outcome $E_{\ell}$
is $P_{\ell}=|c_{\ell}|^{2}$. Repeating the QND measurement will
reproduce the particular $E_{\ell}$ with certainty, with the system
remaining in $\ket{\ell}$. The above discussion is readily extended
to mixed initial system states, and to initial meter states e.g. as
coherent states.

In an analog quantum simulator setting, QND measurement of the many-body
Hamiltonian $\hat{H}$ is incorporated by engineering the extended
system-meter Hamiltonian $\hat{H}_{\mathcal{SM}}=\hat{H}\otimes\mathbb{I}+\vartheta\hat{H}\otimes\hat{P}$.
In an interaction picture with respect to $\hat{H}\otimes\mathbb{I}$,
the joint system then evolves according to the Hamiltonian $\hat{H}_{\mathcal{SM}}^{{\rm int}}\equiv\hat{H}_{{\rm QND}}$
realising the QND measurement discussed above and illustrated in Fig.~\ref{fig:1}(b).
On the other hand, by allowing the system--meter coupling $\vartheta(t)$
to be switched on and off in time, we can alternate between the conventional
free-evolution simulation and QND measurement mode of the system.
In an actual implementation, as discussed below for trapped ions,
we will achieve building the extended system-meter Hamiltonian 
\begin{equation}
\hat{H}_{\mathcal{SM}}=\hat{H}'\otimes\mathbb{I}+\vartheta\hat{H}\otimes\hat{P},\label{eq:H_sys}
\end{equation}
where $\hat{H}'$ and $\hat{H}$ may differ (slightly). We note that the QND
measurement of $\hat{H}$ is obtained by fine-tuning $\smash{\hat{H}'=\hat{H}}$.
A mismatch $\smash{\hat{H}'\neq\hat{H}}$ will be visible as quantum jumps
between energy eigenstates in repeated measurements.

In the trapped-ion example discussed below the many-body Hamiltonian
$\hat{H}$ will be a long-range transverse Ising model~\cite{Zhang2017,Britton2012,Jurcevic:2014aa,Porras2004},
\begin{equation}
\hat{H}=-\sum_{i<j}^{N}J_{ij}\hat{\sigma}_{i}^{x}\hat{\sigma}_{j}^{x}-h\sum_{j}^{N}\hat{\sigma}_{j}^{z},\label{eq:Ising_H}
\end{equation}
where $J_{ij}=J/\abs{i-j}^{\alpha}$ with $\smash{0<\alpha<3}$ and
$h$ the transverse field. Remarkably, in our implementation, the
Hamiltonian $\hat{H}'$ will differ from $\hat{H}$ just by the transverse
field taking on the value $h'$. We will be able to tune $h=h'$ thus
achieving the QND condition.

As last step in our formal development, we wish to formulate QND measurement
of $\hat{H}$ as measurement continuous in time~\cite{Wade2015,Mazzucchi2016,Ashida2017,Lee2014}.
Physically, this amounts to making frequent and, in a continuum limit,
continuous readouts $X(t)$ of the meter variable $\hat{X}$, with
the quantum many-body system evolving according to \eqref{eq:H_sys}.
Following a well-established formalism of quantum optics~\cite{Gardiner2015,Wiseman2010a}
we write for the system under continuous observation a stochastic
master equation (SME) for a conditional density matrix $\hat{\rho}_{\rm c}(t)$
of the many-body system. In our context this SME reads 
\begin{align}
d\hat{\rho}_{\rm c}(t)= & -i[\hat{H}',\hat{\rho}_{\rm c}(t)]dt+\gamma\mathcal{D}[\hat{H}/J]\hat{\rho}_{\rm c}(t)\,dt\nonumber \\
 & +\sqrt{\gamma\epsilon}\mathcal{H}[\hat{H}/J]\hat{\rho}_{\rm c}(t)\,dW(t),\label{eq:SME}\\
dX(t)\equiv & I(t)dt=2\sqrt{\gamma\epsilon}\langle\hat{H}/J\rangle_{\rm c}dt+dW(t).\label{eq:current}
\end{align}
with $dW(t)$ a Wiener increment, to be interpreted as an It{ô} stochastic
differential equation. In a quantum optical setting, as in the ion trap example
below, $I(t)$ is identified with photocurrent in homodyne detection of scattered
light \cite{Gardiner2015}. Monitoring the photocurrent
$\smash{I(t)\sim\langle\hat{H}\rangle_{\rm c}}$ thus provides continuous read out of
the many-body Hamiltonian $\hat{H}$ with
$\left\langle \ldots\right\rangle _{\rm c}\equiv{\rm Tr}[\ldots\hat{\rho}_{\rm c}(t)]$
up to shot noise. Thus $\hat{\rho}_{\rm c}(t)$ describes the many-body quantum state
conditional to observing a particular photocurrent trajectory $I(t)$, as can be
observed in a single run of an experiment.  In \eqref{eq:SME} and
\eqref{eq:current} $\gamma$ is an effective measurement rate, and $\epsilon$ is
a measurement efficiency. Furthermore, we have defined a Lindblad superoperator
$\mathcal{D}[\hat{s}]\hat{\rho}_{\rm c}\equiv\hat{s}\hat{\rho}_{\rm c}\hat{s}^{\dagger}-(\hat{s}^{\dagger}\hat{s}\hat{\rho}_{\rm c}+\mathrm{H.c.})/2$
describing decoherence due to the quantum measurement backaction, and the
nonlinear superoperator
$\mathcal{H}[\hat{s}]\hat{\rho}_{\rm c}\equiv(\hat{s}-\langle\hat{s}\rangle_{\rm c})\hat{\rho}_{\rm c}+\mathrm{H.c.}$
which updates the density matrix conditioned on the observation of the homodyne
photocurrent. Finally, not reading the meter, i.e. averaging over all
measurement outcomes $I(t)$, the SME~\eqref{eq:SME} reduces to a master equation
with Lindblad term $\smash{\sim\mathcal{D}[\hat{H}]\hat{\rho}}$, i.e. realizing
a reservoir coupling with jump operator $\hat{H}$, which erases all off-diagonal
terms of the averaged density matrix $\hat{\rho}$ in the energy eigenbasis.

Equations~\eqref{eq:SME}, \eqref{eq:current} allow us to simulate single
measurement runs corresponding to a stochastic trajectory $I(t)$.
Fig.~\ref{fig:1}(c) illustrates ideal QND measurement,
$\smash{\hat{H}'=\hat{H}}$, of the Hamiltonian~\eqref{eq:Ising_H} by plotting a
sample trajectory of a filtered photocurrent, obtained by averaging $I(t)$ over
a time window $\tau$,
$\mathcal{I}_{\tau}(t)=(2N\sqrt{\gamma\epsilon}\tau)^{-1}\int_{0}^{\infty}dt'I(t-t')e^{-t'/\tau}$.
As initial condition we take all spins pointing against the transverse field. As
seen in Fig.~\ref{fig:1}(c) the trajectory $\mathcal{I}_{\tau}(t)$ (red curve)
stabilizes on a time scale $\sim\gamma^{-1}$ on a particular energy eigenvalue
$E_{\ell}$ of~\eqref{eq:Ising_H} (up to fluctuations from shot noise). In this
figure we consider and show only the eigenstates and eigenenergies~(thin
horizontal lines) within the symmetry sector containing the ground state of the
Ising model with $J,\,h>0$, see Methods. The collapse, and thus preparation of
the many-body wavefunction in the corresponding energy eigenstate is indicated
by plotting the populations
$P_{\ell}(t)\equiv\bra{\ell}\hat{\rho}_{\rm c}(t)\ket{\ell}$ {[}blue shadings in
Fig.~\ref{fig:1}(c){]}. Figure~\ref{fig:1}(d) shows quantum jumps between energy
eigenstates induced by $\hat{H}'\neq\hat{H}$.  For weak perturbation
($\big|[\hat{H}',\hat{H}]\big|\ll|\hat{H}|^{2}$) there are rare jumps between
the energy eigenstates, indicated as $t_{1}$ and $t_{2}$ for the trajectory in
Fig.~\ref{fig:1}(d).  Finally, Fig.~\ref{fig:1}(e) plots the integrated current
$\overline{\mathcal{I}}_{\tau}=(2N\sqrt{\gamma\epsilon}\tau)^{-1}\int_{0}^{\tau}I(t)dt$
and its fluctuations as a function of total integration time $\tau$.  For $N=8$
spins starting in a thermal state the integrated current
$\overline{\mathcal{I}}_{\tau}$ (red curve) exhibits a collapse at a rate
$\smash{\sim\gamma}$ to a particular energy eigenstate. The insets shows the
probabilities $P_{\ell}$ for various times, and the narrowing of the energy
resolution as $\Delta E/J\sim1/\sqrt{\gamma\epsilon\tau}$ with growing $\tau$
(see Methods); first to small energy window containing a few eigenstates as in a
microcanonical ensemble, and eventually to a single energy eigenstate.

\subsection{Implementation with Trapped Ions} 
\begin{figure*}[ht]
\centering \includegraphics[width=1\textwidth]{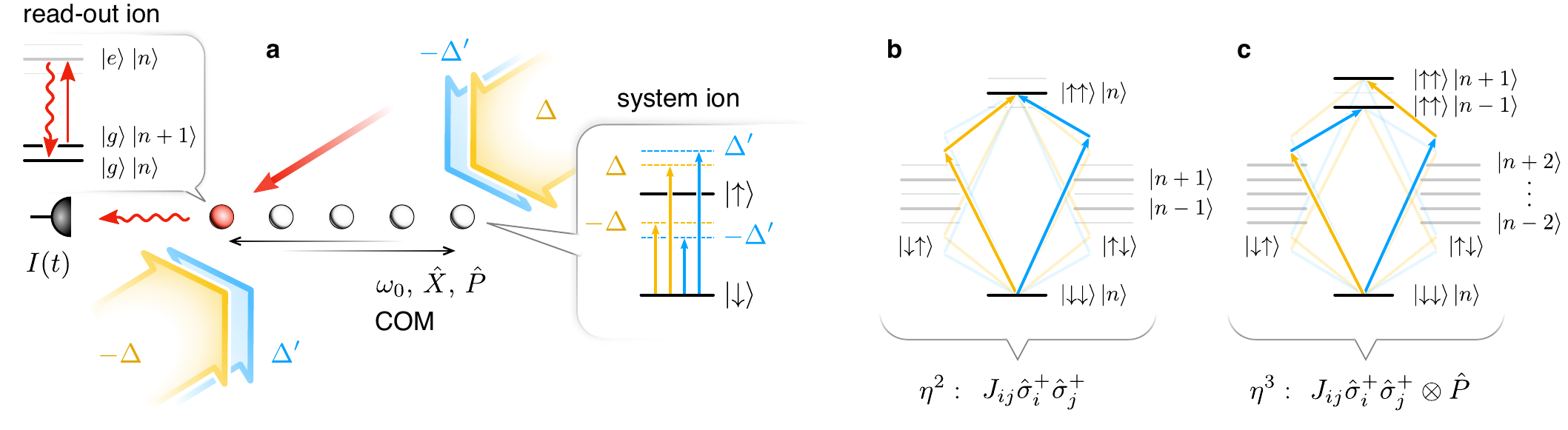} 
\caption{Trapped-ion implementation of the system-meter Hamiltonian $\hat{H}_{\mathcal{SM}}$.
(a) Ion string with $N$ system ions (white) illuminated by four laser
beams in a double M{\o}lmer-S{\o}rensen configuration. As described in the
text this generates $\hat{H}_{\mathcal{SM}}$ {[}see Eq.~\eqref{eq:H_sys}{]}
with transverse Ising Hamiltonians $\hat{H'}$~\eqref{eq:HB} and
$\hat{H}$~\eqref{eq:Ising_H}, and the meter variable $\hat{P}$
representing the COM motion. The meter variable $\hat{X}$ is read
by driving one, or potentially several ancilla ions (red) with a laser
(red beam) tuned to the red motional COM sideband (see text). Homodyne
detection of the scattered light to read $\hat{X}$, and thus revealing
$\hat{H}$ in the photocurrent $I(t)\sim\langle\hat{H}\rangle_{\rm c}$
{[}see Eq.~\eqref{eq:current}{]}. (b) Level scheme of a pair of
ions sharing the COM phonon mode, illustrating one of the elementary
processes contributing to the Ising term $-\sum_{i<j}J_{ij}\hat{\sigma}_{i}^{x}\hat{\sigma}_{j}^{x}\otimes\mathbb{I}$
in second order in $\eta$. (c) Level scheme showing the corresponding
third order processes contributing to $-\vartheta\sum_{i<j}J_{ij}\hat{\sigma}_{i}^{x}\hat{\sigma}_{j}^{x}\otimes\hat{P}$
(see text).}
\label{fig:2}
\end{figure*}
We now provide a trapped ion
implementation of the system-meter Hamiltonian $\hat{H}_{\mathcal{SM}}$
(\ref{eq:H_sys}). As shown in Fig.~\ref{fig:2}(a), we consider a string of $N$
ions in a linear Paul trap representing spin-1/2
$\left\{ \left|\downarrow\right\rangle _{i},\left|\uparrow\right\rangle
  _{i}\right\} $.
These two-level ions can be driven by laser light
$\left|\downarrow\right\rangle \rightarrow\left|\uparrow\right\rangle $, where
the recoil associated with absorption and emission of photons provides a
coupling to vibrational eigenmodes of the ion chain. This includes in particular
the center-of-mass motion (COM) with $\hat{X}$ and $\hat{P}$ position and
momentum operators, respectively, which play the role of meter variables.

To generate in $\hat{H}_{\mathcal{SM}}$ both the Ising interaction
$-\sum_{i<j}J_{ij}\hat{\sigma}_{i}^{x}\hat{\sigma}_{j}^{x}\otimes\mathbb{I}$,
as well as the Ising term coupled to COM, $-\vartheta\sum_{i<j}J_{ij}\hat{\sigma}_{i}^{x}\hat{\sigma}_{j}^{x}\otimes\hat{P}$,
we choose a laser configuration consisting of two pairs of counterpropagating
laser beams {[}c.f. Fig.~\ref{fig:2}(a){]}. In generalization of~\cite{Sorensen1999,Molmer1999}
we call this a double M{\o}lmer-S{\o}rensen configuration. The first pair
of MS beams (shown as amber in Fig.~\ref{fig:2}) is detuned by $\pm\Delta$
from ionic resonance, while the second pair (blue) is detuned by
$\pm\Delta'$. Furthermore, we choose $\Delta'-\Delta=\omega_{0}$
with $\omega_{0}$ the COM frequency. These four laser beams give
rise to laser induced two-photon processes involving pairs of ions,
which are depicted in Figs.~\ref{fig:2}(b,c).

First, as shown in Fig.~\ref{fig:2}(b), absorption of a photon from the one of
the amber MS laser beam followed by an absorption from the counterpropagating
amber beam gives rise to a two-photon excitation
$\ket{\downarrow\downarrow}\rightarrow\ket{\uparrow\uparrow}$, which is resonant
with twice the (bare) ionic transition frequency of the two-level ion. We
emphasize that this process leaves the motional state of the ion chain
unchanged, as illustrated by $\ket{n}\rightarrow\ket{n}$ for the COM mode with
$n$ the phonon occupation number. This process will thus contributes a term
$\sim\hat{\sigma}_{i}^{+}\hat{\sigma}_{j}^{+}$ to the effective spin-spin
interaction. The second pair of MS beams (blue) will again contribute a resonant
two-photon excitation, which adds coherently to the first contribution. By
considering all possible processes, we obtain the effective Ising interaction
$-\sum_{i<j}J_{ij}\hat{\sigma}_{i}^{x}\hat{\sigma}_{j}^{x}\otimes\mathbb{I}$ in
$\hat{H}_{\mathcal{SM}}$. An explicit expression for $J_{ij}$ is given in
Methods in second order perturbation theory in the Lamb-Dicke parameter
$\eta=k/\sqrt{2m\omega_{0}}\ll1$, where $m$ is the ion mass, and $k$ is the
magnitude of the laser wavevector.

Second, with the choice $\Delta'-\Delta=\omega_{0}$ two-photon processes
involving absorption from an amber MS beam and a blue MS beam will be detuned by
the COM frequency from two-photon resonance, i.e. be resonant with the motional
sidebands $\pm\omega_{0}$ {[}c.f.  Fig.~\ref{fig:2}(c){]}. These processes will
change the phonon number by one, and by considering all possible processes
contribute a term
$-\vartheta\sum_{i<j}J_{ij}\hat{\sigma}_{i}^{x}\hat{\sigma}_{j}^{x}\otimes\hat{P}$
to $\hat{H}_{\mathcal{SM}}$. Here $\vartheta\simeq-\eta\sqrt{2/N}$, and $J_{ij}$
is identical to the couplings obtained above. We note that this term is of order
$\eta^{3}$ (for details see Methods).

By considering a (small) imbalance of Rabi frequencies in MS laser
configurations, we can create in $\hat{H}_{\mathcal{SM}}$ a transverse-field
term $-h\sum_{j}\hat{\sigma}_{j}^{z}\otimes\mathbb{I}$, and in addition a term
$+\vartheta h\sum_{j}\hat{\sigma}_{j}^{z}\otimes\hat{P}$ (see Methods and
Supplementary Note~\ref{sec:implementation-ions}). Thus, our laser configuration generates $\hat{H}$ and $\hat{H}'$ with the
same Ising term but opposite transverse field $\pm h$. To rectify the
transverse-field mismatch, we can offset the detuning of the four lasers by a
small amount $\pm\Delta^{(\prime)}\to\pm\Delta^{(\prime)}-2B$.  We obtain
$\hat{H}$ as in Eq.~\eqref{eq:Ising_H} and
\begin{equation}
\hat{H}'=-\sum_{i<j}^{N}J_{ij}\hat{\sigma}_{i}^{x}\hat{\sigma}_{j}^{x}-(B-h)\sum_{j}^{N}\hat{\sigma}_{j}^{z}.\label{eq:HB}
\end{equation}
The choice $B=2h$ thus allows us to tune to the QND sweetspot $\hat{H}'=\hat{H}$
as in Fig.~\ref{fig:1}(c,e), while away from this point we obtain
$\hat{H}'\ne\hat{H}$ as considered in Fig.~\ref{fig:1}(d).

Finally, the homodyne current~\eqref{eq:current} corresponding to
a continuous measurement of the COM quadrature $\hat{X}$, and thus
of the Hamiltonian $\hat{H}$ can be measured via homodyne detection
of the scattered light from an ancillary ion driven by a laser on
the red motional COM sideband {[}c.f. Fig.~\ref{fig:2}(a) and Methods{]}.

\begin{figure*}[ht!]
\includegraphics[width=1\textwidth]{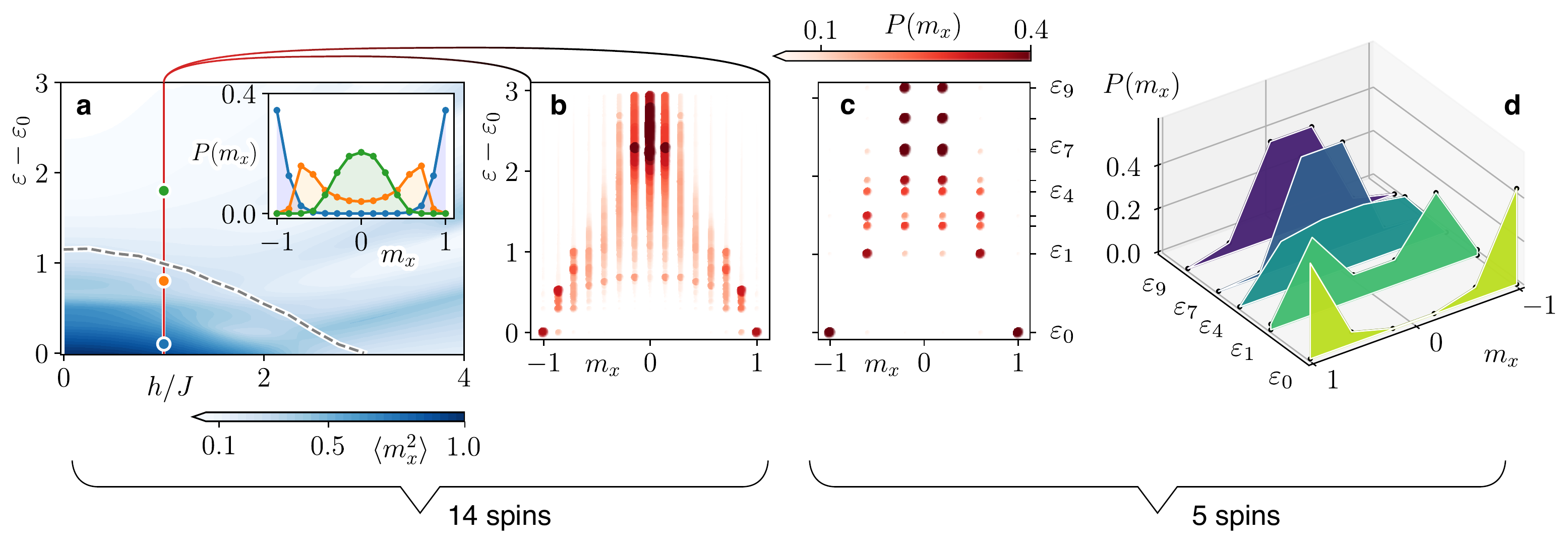}

\caption{Excited-state phase transition in the Ising model~\eqref{eq:Ising_H}
with $\alpha=1.5$. (a) Ferro-paramagnet crossover in the Ising model
of $N=14$ spins prepared by the energy measurements in microcanonical
ensembles of width $\Delta E/(JN)=0.1$. The transition between magnetically
ordered phase $\braket{\hat{m}_{x}^{2}}_{\text{mc}}\approx1$ (dark
blue) to disordered phase $\braket{\hat{m}_{x}^{2}}_{\text{mc}}\approx0$
(light blue) is shown as function of the mean energy density $\varepsilon=\braket{\hat{H}}_{{\rm mc}}/(JN)$
and the transverse field $h$. An estimate of the critical energy
density in the thermodynamic limit, obtained with Monte-Carlo simulation
of canonical ensembles of $512$ spins with rescaled interactions
(see Methods and Supplementary Note~\ref{sec:numer-study-therm}), is shown as black dashed line. The inset shows the
order parameter distribution $P\left(m_{x}\right)$ for $h/J=1$ and
$\varepsilon=0.1,\,0.8,\,1.8$ in blue, orange, and green, respectively.
Test of ETH (within the symmetry sector $\left\{ +1,-1\right\} $
see Methods): (b) order-disorder transition is seen as crossover from
bi-modal distribution of $P\left(m_{x}\right)$ at low energies to
a single-peak distribution at high energies, shown on the level individual
eigenstates. Color intensity and the dot size indicate the corresponding
probability. (c, d) Qualitatively similar energy dependence of $P\left(m_{x}\right)$
shown for a system of just 5 spins. (d) Signatures of the phase transition
visible for representative sample of eigenstates.}
\label{fig:3}
\end{figure*}

\subsection{QND measurement protocols} Implementation of
$\hat{H}_{\mathcal{SM}}$ with time-dependent system-meter coupling
$\vartheta(t)$ allows protocols where we switch between time-windows of
unobserved quantum simulation, and measurement of energy, and thus preparation
of energy eigenstates, which is verified by observing convergence of the
filtered photocurrent.  In addition, the Hamiltonians~\eqref{eq:Ising_H}
and~\eqref{eq:HB} can be made time-dependent, e.g. with a time-dependent
magnetic field.  This allows us to perform work on the system, and measure work
distribution functions via measurement of energy~\cite{Campisi2011}. Our QND
toolbox thus opens up the door to address experimentally fundamental problems of
(non-equilibrium) statistical mechanics in analog quantum simulation. We apply
the QND toolbox below first to ETH~\cite{DAlessio2016,Deutsch2018} and then we
consider testing QFRs~\cite{Campisi2011} in interacting many-body systems. We
emphasize that our setting explores naturally the interesting regime of
mesoscopic particle numbers from a few to tens of spins.

\begin{figure*}
\includegraphics[width=0.9\textwidth]{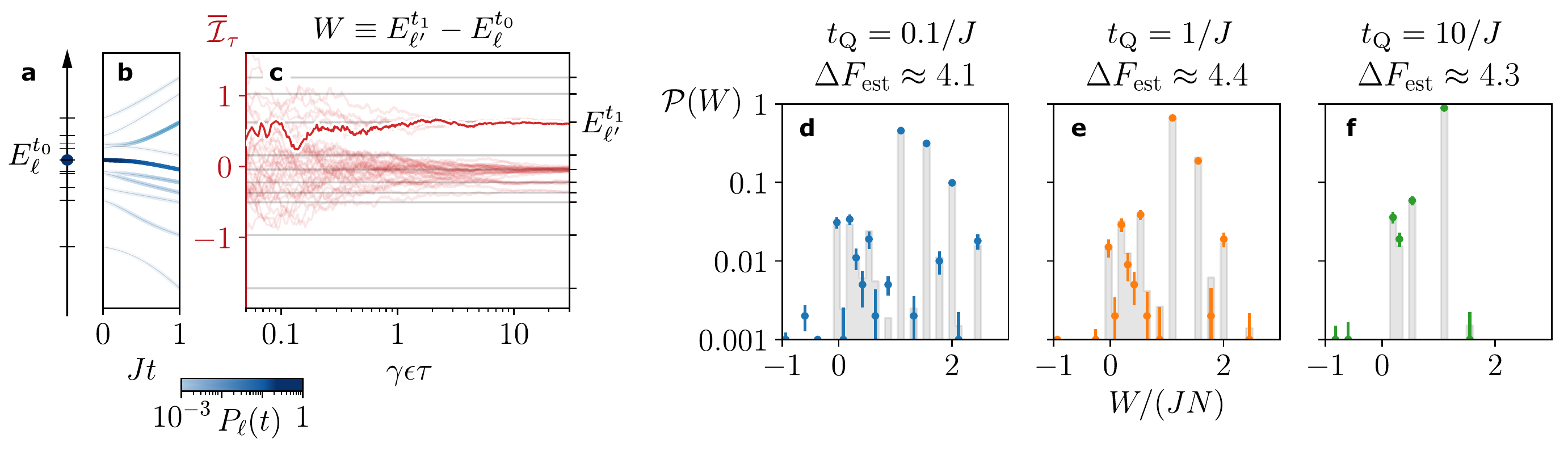}
\caption{Verification of Jarzynski equality~\eqref{eq:Jarzynski} for the transverse field Ising model with 5 spins.
A single realisation of the proposed protocol consists of (a) the projection of the initial thermal state of spins to an eigenstate $E_{\ell}^{t_0}$ of the initial Hamiltonian with transverse field value $h_{t_0}$, (b) the free evolution under a linear change of the transverse field $h$ during time $t_{\rm Q}=t_1-t_0$ (probabilities $P_{\ell}(t)$ to populate the instantaneous energy eigenstates are shown in shades of blue), (c) the final energy readout $E_{\ell}^{t_1}$ of the system at transverse field $h_{t_1}$ (photocurrent realizations for simulated experimental runs are shown in red) resulting in the work $W$ (see text) performed during the non-equilibrium process.
(d,e,f)~The resulting work probability distribution $\mathcal{P}(W)$ for $h_{t_0}=2J$, $h_{t_1}=0.5J$, $\alpha=2$, $\beta=0.5/J$, and various quench times $t_{\rm Q}$. The gray bars show theoretical probability as a function of work $W$. The color dots with the corresponding vertical error bars (one standard deviation) show the estimated probabilities for 1000 simulated experimental runs. Independent of the quench duration the Jarzynski relation~\eqref{eq:Jarzynski} yields similar estimations of the free energy
difference $\Delta F_{\rm est}$ (fluctuating due to the finite number of runs) with the true value given by $\Delta F\approx4.35$.}
\label{fig:4} 
\end{figure*}

\subsection{Thermal properties of energy eigenstates} Single energy eigenstates
$\ket{\ell}$ can encode thermal properties which we typically associate with a
microcanonical or canonical ensemble describing systems in thermodynamic
equilibrium. This eigenstate thermalization concerns on the one hand expectation
values of few-body observables, leading to the remarkable prediction of the ETH
that diagonal matrix elements $\braket{\ell|\hat{O}|\ell}$ have to agree with
the microcanonical average at energy $E_{\ell}$,
$\braket{\ell|\hat{O}|\ell}=O(E_{\ell})=\tr(\hat{O}\hat{\rho}_{E_{\ell}}^{\mathrm{mc}})$.
Here $\hat{\rho}_{E_{\ell}}^{\mathrm{mc}}$ is the microcanonical density
operator as a mixture of energy eigenstates within a narrow range centered
around $E_{\ell}$. On the other hand, ETH imposes constraints on dynamical
properties for diagonal and off-diagonal matrix elements
$\braket{\ell'|\hat{O}|\ell}$; e.g. two-time correlation functions and dynamical
susceptibilities have to be related by the fluctuation-dissipation
theorem~\cite{DAlessio2016}. To be more specific, ETH suggests a
structure~\cite{Srednicki1999}
$\braket{\ell'|\hat{O}|\ell}=O(\bar{E})\delta_{\ell'\ell}+e^{-S(\bar{E})/2}f_{\hat{O}}(\bar{E},\omega)R_{\ell'\ell}$
where diagonal and off-diagonal matrix elements are determined by the functions
$O(\bar{E})$ and $f_{\hat{O}}(\bar{E},\omega)$, respectively, which depend
smoothly on their arguments $\bar{E}=(E_{\ell}+E_{\ell'})/2$ and
$\omega=E_{\ell'}-E_{\ell}$. $S(\bar{E})$ is the thermodynamic entropy at the
mean energy $\bar{E}$, and $R_{\ell'\ell}$ is a random number with zero mean and
unit variance. An experimental test of ETH, therefore, requires the ability to
measure both diagonal and off-diagonal elements, something which is provided by
our ion toolbox.

The transverse Ising model~\eqref{eq:Ising_H}, as realized with ions, provides a
rich testbed for ETH~\cite{Fratus2016}. For $1<\alpha\leq2$, this model features
a ferromagnetic transition at finite temperature or energy density, in a
canonical or microcanonical description, respectively. As illustrated above in
Fig.~\ref{fig:1}, our trapped-ion QND toolbox enables the preparation of
microcanonical ensembles of variable width $\Delta E$. According to ETH, the
ferromagnetic transition persists even in the limit of vanishing $\Delta E$,
which corresponds to the preparation of a single energy eigenstate.

For reference, the microcanonical phase diagram at finite $\Delta E$ is shown in
Fig.~\ref{fig:3}(a) for an experimentally accessible system size of $N=14$ spins
and $\alpha=1.5$ (see Supplementary Note~\ref{sec:implementation-ions} for experimental parameters). The
ferromagnetic transition is clearly manifest in the distribution $P(m_{x})$ of
the magnetization $\hat{m}_{x}=N^{-1}\sum_{j}\hat{\sigma}_{j}^{x}$, which is
bimodal in the ferromagnetic phase, see the inset in Fig.~\ref{fig:3}(a).
Consequently, fluctuations $\braket{\hat{m}_{x}^{2}}$ are finite (vanish) in the
ferromagnetic (paramagnetic) phase, and indicate order even in the absence of
symmetry-breaking fields. A trapped-ion quantum simulator provides the ability
to perform single-site resolved read-out of spins, thus giving direct access to
the distribution $P(m_{x})$ and, consequently, the fluctuations
$\braket{\hat{m}_{x}^{2}}$. Due to the quasi-diagonal structure of
$\braket{\ell'|\hat{O}|\ell}$ for ETH-satisfying observables $\hat{O}$, the
hypothesis is expected to hold for any power of such observables and, in
particular, also for the full probability distribution function
$P(m_{x})$~\cite{Srednicki1999}. Indeed, as we show in Figs.~\ref{fig:3}(b)
and~(c,d), respectively, we find clear signatures of the transition in
$P(m_{x})$ for individual energy eigenstates both for $N=14$ and even much
smaller system of only $N=5$ spins, in which single eigenenergies can be
resolved with current experimental technology (see Supplementary Note~\ref{sec:implementation-ions}).

The observation of the Ising transition in single eigenstates gives a
qualitative indication of eigenstate thermalization in diagonal matrix
elements. A stringent quantitative assessment requires to show that fluctuations
of single-eigenstate expectation values $\braket{\ell|\hat{O}|\ell}$ around the
microcanonical average $\tr(\hat{O}\hat{\rho}_{E_{\ell}}^{\mathrm{mc}})$ are
suppressed with increasing system size~\cite{DAlessio2016}. We discuss
experimental requirements for such a test, along with a protocol to measure
off-diagonal matrix elements, in Supplementary Note~\ref{sec:eigenst-therm-hypoth}.

\subsection{Work distribution function and quantum fluctuation relations}
Projective measurements in the energy eigenbasis are the key ingredient for the
long-sought experimental verification of QFRs~\cite{Campisi2011}. The
challenging requirement to measure changes in the energy of the system on the
level of single energy eigenstates has been achieved only recently in
single-particle systems~\cite{Batalhao2014,Zhang2014}. Our QND measurement
scheme opens up the possibility to probe QFRs in a true many-body setting.

As an illustration we consider the celebrated Jarzynski equality which describes the mean value of the exponentiated work performed on a system in an arbitrary non-equilibrium process defined by a time-dependent Hamiltonian $\hat{H}(t)$~\cite{Campisi2011}. The equality relates the work to the difference between the free energies $\Delta F$ of equilibrium systems described by the Hamiltonian at the initial $t_0$ and the final $t_1$ times:
\begin{equation}
\left\langle e^{-\beta W}\right\rangle = e^{-\beta \Delta F}.\label{eq:Jarzynski}
\end{equation}
Here $\beta$ is the inverse temperature specifying an initial canonical thermal state of the system $\hat{\rho}_{\beta}=e^{-\beta \hat H(t_0)}/Z_{t_0}$
with $Z_{t_0}=\tr[e^{-\beta \hat{H}(t_0)}]$ and $\Delta F = F_{t_1}-F_{t_0} = -\ln(Z_{t_1}/Z_{t_0})/\beta$. The average on the left hand side of Eq.~\eqref{eq:Jarzynski} is performed with respect to the distribution of work $\mathcal{P}(W)$ (see Methods). The work itself is determined as the difference between the outcomes of two energy measurements before and after the time-dependent protocol $W\equiv E_{\ell'}^{t_1}-E_{\ell}^{t_0}$. Remarkably, while the work distribution does depend on details of the time evolution given by $\hat{H}(t)$, the average is defined only by the initial and final Hamiltonians. Therefore, the QFR enables experimental measurements of the equilibrium property $\Delta F$ via measurement of work in a non-equilibrium process.

Our scheme provides the required ingredients for probing the QFR in the
interesting regime of intermediate system sizes which is dominated by quantum
fluctuations: As presented above, the scheme allows for independent temporal
control of parameters of the spin Hamiltonian as well as the system-meter
coupling in Eq.~\eqref{eq:H_sys}. Further, single energy levels are well
resolved for a system of five interacting spins as we consider in the following
(see Supplementary Note~\ref{sec:implementation-ions} for experimental parameters).

This enables the protocol shown in Fig.~\ref{fig:4}(a-c), in which a first
measurement of the energy is carried out while the magnetic field $h$ in
Eq.~\eqref{eq:Ising_H} is kept constant at a value $h_{t_0}$.  Then, the
system-meter coupling is switched off, and the magnetic field is linearly ramped
to a final value $h_{t_1}$, followed by another measurement. The statistics of
corresponding measurement outcomes, $E_{\ell}^{t_0}$ and $E_{\ell'}^{t_1}$,
determines the distribution of work $\mathcal{P}(W)$ performed on the system
during the magnetic field ramp. Initialization of the system at arbitrary
temperatures can be emulated by weighting different runs of the protocol
according to the Gibbs distribution $e^{-\beta E_{\ell}^{t_0}}/Z_{t_0}$ with the
initial energy $E_{\ell}^{t_0}$.

The resulting work distributions $\mathcal{P}(W)$ for various quench durations
$t_{\rm Q}$ are shown in Fig.~\ref{fig:4}(d-f). While the probability distribution for
fast (blue dots) and slow (green dots) quenches differs significantly the
estimated free energy difference
$\Delta F_{\rm est} = -\ln\langle e^{-\beta W}\rangle/\beta$ approximately (due
to the finite number of simulated experimental runs) matches the true value of
$\Delta F$ for all three quench speeds, thus, verifying the
equality~\eqref{eq:Jarzynski}.

\section{Discussion}
We have developed a QND toolbox in analog quantum simulation realising
single-shot measurement of the energy of an isolated quantum many-body
system, as a key element towards experimental studies in non-equilibrium
quantum statistical mechanics. This comprises ETH and quantum thermodynamics,
including quantum work distribution, and Jarzynski and Crooks fluctuations
relations~\cite{Campisi2011} in mesoscopic quantum many-body systems. The present
work outlines an ion-trap implementation with COM phonons as meter.
However, the concepts and techniques carry over to other platforms
including CQED with atoms~\cite{RevModPhys.85.553} and superconducting
qubits~\cite{Houck:2012aa}, where the role of the meter can be represented
by cavity photons read with homodyne detection, and Rydberg tweezer
arrays~\cite{Norcia2018,Cooper2018,Keesling:2019aa,Barredo:2018aa,PhysRevLett.122.143002}
by coupling to a small atomic ensemble encoding the continuous meter
variables~\cite{hammerer2010quantum}, respectively. Finally, while
the present work considers QND measurement of the total Hamiltonian
$\hat{H}$ of an isolated system, our approach generalizes to measuring
Hamiltonians $\hat{H}_{A}$ of subsystems, as is of interested in
quantum transport of energy, or energy exchange in coupling the many-body
system of interest to a bath.

\section{Methods}

\subsection{System-meter coupling Hamiltonian}
We choose for the four
lasers in our double MS configuration the detuning and the Rabi frequency
as $(\Delta,\Omega)$, $(-\Delta,\Omega+\delta\Omega)$, $(\Delta',\Omega)$,
$(-\Delta',\Omega+\delta\Omega)$, where $\delta\Omega\propto\eta^{2}\Omega$
is a small imbalance we use to generate the transverse field term
in the spin model. We are interested in the regime of sufficiently
large detunings compared to the Rabi frequency $\Omega$, such that
single lasers only virtually excite the ions and the phonon modes,
$\Omega\ll\Delta^{(\prime)}$, $\eta_{q}\Omega\ll|\Delta^{(\prime)}-\omega_{q}|$,
where $\omega_{q}$ is the oscillation frequency of the $q$-th phonon
mode and $\eta_{q}\equiv\eta\sqrt{\omega_{0}/\omega_{q}}$. On large
timescales $t\gg1/\Delta^{(\prime)},1/\omega_{0}$, we obtain an effective
Hamiltonian $\hat{H}_{{\cal SM}}$ describing the coupled dynamics
of the system and the meter, i.e., the spins and the COM phonon mode,
by performing the Magnus expansion~\cite{BLANES2009151} to the time
evolution operator in the interaction picture (see Supplementary Note~\ref{sec:implementation-ions}). We
further expand $\hat{H}_{{\cal SM}}$ in terms of $\eta$. In second
order in $\eta$ we recover the transverse field Ising Hamiltonian~\cite{Zhang2017,Britton2012,Jurcevic:2014aa},
$\hat{H}_{{\cal SM}}^{\left(2\right)}=(-\sum_{i<j}J_{ij}\hat{\sigma}_{i}^{x}\hat{\sigma}_{j}^{x}+h\sum_{j}\hat{\sigma}_{j}^{z})\otimes\mathbb{I}\equiv\hat{H}'\otimes\mathbb{I}$,
where the spin-spin couplings 
\begin{equation}
J_{ij}=-\eta^{2}\omega_{0}\sum_{q}M_{iq}M_{jq}\left[\frac{\Omega^{2}}{\Delta^{2}-\omega_{q}^{2}}+\frac{\Omega^{2}}{(\Delta^{\prime})^{2}-\omega_{q}^{2}}\right]
\end{equation}
include contributions from the two MS configurations independently
with $M_{iq}$ denoting the distribution matrix element of the $q$-th
phonon mode. The transverse field strength is $h=\Omega\delta\Omega\left(1/\Delta+1/\Delta^{\prime}\right)/2$.

Crucially, under the condition $\Delta'-\Delta=\omega_{0}$, the crosstalk
between the two MS configurations leads to an extra resonant processes
as exemplified by Fig.~\ref{fig:2}(c). These are described by expanding
the effective Hamiltonian $\hat{H}_{{\cal SM}}$ to third order in
$\eta$, 
$\hat{H}_{{\cal SM}}^{\left(3\right)}=(-\eta\sqrt{2}M_{i0})(-\sum_{i<j}J_{ij}\hat{\sigma}_{i}^{x}\hat{\sigma}_{j}^{x}-h\sum_{j}\hat{\sigma}_{j}^{z})\otimes\hat{P}\equiv{\vartheta}\hat{H}\otimes\hat{P}$,
where $M_{i0}\simeq1/\sqrt{{N}}$ is the (equal) distribution matrix
element of the COM mode. Combining $\hat{H}_{{\cal SM}}^{\left(2\right)}$
and $\hat{H}_{{\cal SM}}^{\left(3\right)}$ gives the desired system-meter
Hamiltonian~\eqref{eq:H_sys}. The transverse field in $\hat{H}_{{\cal SM}}^{\left(2\right)}$
and $\hat{H}_{{\cal SM}}^{\left(3\right)}$ can be independently tuned
with the method discussed in the main text. Higher order terms beyond
$\hat{H}_{{\cal SM}}^{\left(3\right)}$ have negligible effects, for
details see Supplementary Note~\ref{sec:implementation-ions}.

Our double MS configuration can be implemented with both axial and
transverse phonon modes. The implementation with transverse modes gives rise to power-law spin interactions $J_{ij}=J/|i-j|^\alpha$ with $0\leq \alpha\leq 3$, which are considered in the rest of this paper.
Experimental considerations and scalability
are discussed in the Supplementary Note~\ref{sec:implementation-ions}.

\subsection{Continuous readout of $\hat{X}$} We assume that in Fig.
\ref{fig:2} the ancillary ion does not see the four MS lasers (amber
and blue) and, similarly, the system ions do not couple to the read-out
laser (red), i.e we assume single-ion addressability~\cite{Linke3305}
or with mixed-species~\cite{Negnevitsky:2018aa}. The read-out laser is tuned
in resonance with the red sideband of the COM mode, $\Delta_{\rm e}=\omega_{0}$,
under the resolved-sideband condition $\omega_{0}\gg\Gamma_{\rm e},\Omega_{0}$,
where $\Gamma_{\rm e}$ is the spontaneous emission rate of the cooling
transition $\left|e\right\rangle \to\left|g\right\rangle $ while
$\Omega_{0}$ and $\Delta_{\rm e}$ are the Rabi frequency and the detuning
of the cooling laser respectively. In this regime, the emitted electric
field is proportional to $\langle\hat{a}_{0}\rangle$ with $\hat{a}_{0}$
the annihilation operator of the COM mode. Homodyne detection
then directly reveals the quadrature of the COM phonon (the meter).
The homodyne current can be written as (see Supplementary Note~\ref{sec:implementation-ions}) 
\begin{equation}
dX(t)\equiv I(t)dt=\sqrt{2\epsilon\gamma_{\rm s}}\langle\hat{X}\rangle_{\rm c}+dW(t),
\end{equation}
where $\epsilon$ is the photon detection efficiency, $\gamma_{\rm s}\simeq k_{0}^{2}\Omega_{0}^{2}/(2\Gamma_{\rm e}Nm_{0}\omega_{0})$
is the measurement rate with $k_{0}$ the cooling laser wavevector
and $m_{0}$ the ancillary ion mass, and we have chosen the frequency and phase of the homodyne local oscillator to maximize the homodyne current. Correspondingly,
the evolution of the conditional state $\rho_{\rm c}^{\mathcal{SM}}(t)$
of spin system plus the meter is described by a SME 
\begin{align}
d\hat{\rho}_{\rm c}^{\mathcal{SM}}(t)= & -i[\hat{H}_{{\cal SM}},\hat{\rho}_{\rm c}^{\mathcal{SM}}(t)]dt+\gamma_{\rm s}\mathcal{D}\left[\hat{a}_{0}\right]\hat{\rho}_{\rm c}^{\mathcal{SM}}(t)dt\nonumber \\
 & +\sqrt{\epsilon\gamma_{\rm s}}\mathcal{H}\left[\hat{a}_{0}\right]\hat{\rho}_{\rm c}^{\mathcal{SM}}(t)dW(t),\label{eq:SME_eff}
\end{align}
Eliminating the meter under the condition $\gamma_{\rm s}\gg|\vartheta J|$,
we realize continuous QND readout of the spin Hamiltonian as described
by Eqs.~\eqref{eq:SME} and ~\eqref{eq:current} with $\gamma=2(\vartheta J)^{2}/\gamma_{\rm s}$.

We further emphasize that the readout laser, which is tuned to the
red sideband, also acts as cooling of the COM mode. Furthermore, the
readout signal can be enhanced with several ancilla ions.

\subsection{Energy measurement resolution} Here we estimate the Signal-to-Noise
Ratio (SNR) which allows us to distinguish two adjacent energy levels
separated by $\Delta E$. The difference of photocurrents~\eqref{eq:current}
corresponding to the two energy levels integrated over time $\tau$
reads 
\[
\int_{0}^{\tau}\!\![I_{1}(t)-I_{2}(t)]dt=\underset{{\rm Signal}}{\underbrace{\vphantom{\int}2\sqrt{\gamma\epsilon}(\Delta E/J)\tau}}+\underset{{\rm Noise}}{\underbrace{\int_{0}^{\tau}\!\![dW_{1}(t)-dW_{2}(t)]}}.
\]
Considering the shot noises $W_{1,2}(t)$ of two measurements as uncorrelated
and using the Wiener increment property $dW_{1,2}^{2}(t)=dt$ we obtain
${\rm SNR}=2\gamma\epsilon(\Delta E/J)^{2}\tau$. For a given averaging
time $\tau$, the condition ${\rm SNR}\gg1$ provides us with the
minimal energy difference we can distinguish $\Delta E/J\gg1/\sqrt{2\gamma\epsilon\tau}$.

\subsection{Symmetries of the long-range transverse field Ising model}
The transverse field Ising model~\eqref{eq:Ising_H} is invariant
under the reflection and spin inversion symmetry transformations.
We now provide an operational definition of these symmetries and the
corresponding symmetry sectors.

Consider a product state vector in the $\sigma^{x}$ basis $\left|\phi\right\rangle =\left|s_{1}^{x}\ldots s_{N}^{x}\right\rangle $.
The reflection operator can be defined by its action on the $\left|\phi\right\rangle $
state as $R\left|s_{1}^{x},\ldots,s_{N}^{x}\right\rangle \equiv\left|s_{N}^{x},\ldots,s_{1}^{x}\right\rangle $.
Analogously, the spin inversion operator can be defined as $P\left|s_{1}^{x}\ldots s_{N}^{x}\right\rangle \equiv\left|-s_{1}^{x},\ldots,-s_{N}^{x}\right\rangle $.
Both operators have two eigenvalues $\pm1$ and commute with each
other and the Hamiltonian Eq.~\eqref{eq:Ising_H}, thus, representing
QND observables which can also be measured in the non-destructive
way as presented in the paper.

The Hamiltonian can be independently diagonalized in each of the subspaces
corresponding to eigenvalues of the $R$ and $P$ operators. The ground
state of the Ising model with $J,\,h>0$ belongs to the $\left\{ +1,+1\right\} $
symmetry sector. For the test of ETH in Fig.~\ref{fig:3} we consider
the symmetry sector with eigenvalues of $R$ and $P$ given by $\left\{ +1,-1\right\} $,
respectively. The subspace can be reached from the $\left\{ +1,+1\right\} $
sector by flipping odd number of spins (along the $\sigma^{z}$ direction)
in the limit of strong transverse field.

\subsection{Interaction renormalization} In numerical simulations in
Fig.~\ref{fig:3} we renormalize the interaction strength coefficient
$J$ such that the average interaction strength matches its value
in thermodynamic limit. More precisely, for the $N$-spin Ising model~\eqref{eq:Ising_H}
we rescale $J\rightarrow J_{N}\equiv J\cdot S_{N}/S_{\infty}$ with
$S_{N}\equiv\frac{1}{N}\sum_{i,j=1}^{N}1/\left|i-j\right|^{\alpha}$.
The results are then expressed in units of $J_{14}$.

\subsection{Work distribution function} The work distribution of a
process defined by a time-dependent Hamiltonian $\hat{H}(t)$ (with
the corresponding instantaneous energy eigenvalues and eigenstates
$E_{\ell}^{t}$ and eigenstates $\psi_{\ell}^{t}$ ) is defined as
follows \cite{Campisi2011} : 
\begin{equation}
\mathcal{P}(W)=\sum_{\ell\ell^{\prime}}\delta[W-(E_{\ell^{\prime}}^{t}-E_{\ell}^{0})]P_{\ell^{\prime}\ell}^{t}P_{\ell}^{0},\label{eq:work_dist}
\end{equation}
where $P_{\ell}^{0}=\left\langle \psi_{\ell}^{0}\right|\rho_{{\rm in}}\left|\psi_{\ell}^{0}\right\rangle $
are the occupation probabilities of the initial state and $P_{\ell^{\prime}\ell}^{t}=|\left\langle \psi_{\ell^{\prime}}^{t}\right|U(t,0)\left|\psi_{\ell}^{0}\right\rangle |^{2}$
are the transition probabilities between initial $\ell$ and final
$\ell^{\prime}$ states with $U(t,0)\equiv\mathcal{T}e^{-i\int_{0}^{t}\hat{H}(t')dt'}$
the evolution operator. The average of the exponentiated work in Eq.
(\ref{eq:Jarzynski}) is readily defined as an integral with the work
distribution function Eq. (\ref{eq:work_dist}): $\langle e^{-\beta
  W}\rangle\equiv\int dWe^{-\beta W}\mathcal{P}(W)$.

\section{Data availability}
The data sets generated and analyzed in the current study are available from the corresponding author upon reasonable request.

\section{Code availability}
The codes used to generate the numerical data in the current study are available from the corresponding author upon reasonable request.

\section{Acknowledgments} We thank M. Rigol, A. S{\o}rensen, M. Srednicki, and
P. Talkner for valuable comments. This work is supported by the European Union’s Horizon 2020 program under Grants Agreement No. 817482 (PASQuanS) and No. 731473 (QuantERA via QTFLAG), the US Air Force Office of Scientific Research (AFOSR) via IOE Grant No. FA9550-19-1-7044 LASCEM, the Austrian Research Promotion Agency (FFG) via QFTE project AutomatiQ, by the Simons Collaboration on Ultra-Quantum Matter, which is a grant from the Simons Foundation (651440, P.Z.). DY acknowledges the financial support by Industriellenvereinigung Tirol. PZ thanks KITP for hospitality as member of the
QSIM19 program, and support through Grant NSF PHY-1748958.  The stochastic
master equation is solved using the open-source \mbox{QuTiP}
package~\cite{JOHANSSON2013}. We use \mbox{QuSpin} for the exact diagonalization
of the Ising model~\cite{Weinberg_17}. For the quantum Monte-Carlo
simulations we use the ALPS code~\cite{Bauer_2011}.

\section{Author contributions}
D.Y., A.G., L.M.S., and D.V.V. designed the model, developed the methods, performed the calculations, and analysed the data. All authors discussed the results and wrote the manuscript. P.Z. conceived the study and was in charge of the overall direction and planning.

\section{Competing interests}
The authors declare no competing interests.

\section{Additional information}
Supplementary information is available for this paper.

\section*{}
\newpage
\begin{center}
\textbf{\large Supplementary Information}
\end{center}

This Supplementary Information is organized as follows:
Supplementary Note~\ref{sec:implementation-ions} contains a detailed discussion of the
implementation of the QND measurement scheme in a trapped-ion quantum simulator,
including an analysis of the experimental feasibility of the scheme with
different species of ions and using transverse or axial phonon modes. We provide
additional information on applications of the QND scheme in
Supplementary Note~\ref{sec:numer-study-therm} and~\ref{sec:eigenst-therm-hypoth}. In
particular, we give details on the numerical analysis of the ferromagnetic
transition in the transverse-field Ising, and discuss prospects of an
experimental test of the ETH.

\section{Implementation with trapped ions}
\label{sec:implementation-ions}



In the main text we outline the implementation of our QND measurement scheme in
a trapped-ion quantum simulator. In this section we elaborate on the detailed
derivations behind the short presentation in the main text, and discuss the
experimental feasibility of the proposed scheme.  The section is structured as
follows.

In Supplementary Note~\ref{subsec:Double-Molmer-Sorensen-interaction} we introduce and provide an analytical study of 
the double M\o lmer-S\o rensen (MS) laser configuration (see Fig.~2 of
the main text). We show that the low-frequency dynamics is governed by the effective system-meter coupling Hamiltonian
$\hat{H}_{{\cal SM}}$, defined in Eq.~(2) of the main text (we set
$\hbar=1$ hereafter) 
\begin{equation}
\hat{H}_{{\cal SM}}=\hat{H}^{\prime}\otimes\mathbb{I}+\vartheta\hat{H}\otimes\hat{P},\label{eq:H_QND}
\end{equation}
where $\hat{P}\equiv i(\hat{a}_{0}^{\dagger}-\hat{a}_{0})/\sqrt{2}$
is the quadrature operator of the center-of-mass (COM) phonon mode,
with $\hat{a}_{0}(\hat{a}_{0}^{\dag})$ the corresponding annihilation(creation)
operator. Both $\hat{H}$ and $\hat{H}^{\prime}$
are many-body spin Hamiltonians of the Ising type, 
\begin{align}
\hat{H} & =-\sum_{i<j}^{N}J_{ij}\hat{\sigma}_{i}^{x}\hat{\sigma}_{j}^{x}-h\sum_{j=1}^{N}\hat{\sigma}_{j}^{z},\label{eq:H_Ising_coupling_to_meter}\\
\hat{H}' & =-\text{\ensuremath{\sum_{i<j}^{N}J_{ij}\hat{\sigma}_{i}^{x}\hat{\sigma}_{j}^{x}}}-(B-h)\sum_{j=1}^{N}\hat{\sigma}_{j}^{z}.\label{eq:H_Ising_intro}
\end{align}
By adjusting the transverse field strength $B$, we are able to tune
the measurement from QND $(B=2h)$ to imperfect QND $(B\simeq2h)$
which supports the observation of quantum jumps.

In Supplementary Note~\ref{sec: Numerical} we provide a numerical study of the double MS scheme in different parameter regimes which 
supports the validity of the system-meter coupling Hamiltonian Eq.~\eqref{eq:H_QND}. 

In Supplementary Note~\ref{sec:continuous_readout} we describe the continuous readout of the
spin Hamiltonian $\hat{H}$, achieved by sideband laser cooling of the motion of
an ancilla ion at the edge of the ion chain and homodyne detection of its
fluorescence, as schematically shown in Fig.~2(a) of the main text. We derive
the resulting dynamics of the spin system as described by the stochastic master
equation (SME)
\begin{eqnarray}
d\hat{\rho}_{c}(t) &=&-i[\hat{{H}}',\hat{\rho}_{c}(t)]dt+\gamma\mathcal{D}[\hat{H}/J]\hat{\rho}_{c}(t)dt \nonumber 
\\
&&+\sqrt{\epsilon\gamma}\mathcal{H}[\hat{H}/J]\hat{\rho}_{c}(t) dW(t).\label{eq:SME_measureH}
\end{eqnarray}
Here $\hat{\rho}_c(t)$ is the density matrix of the spin system conditioned on the homodyne detection signal, $J$ is the characteristic energy scale of the spin Hamiltonian, $\gamma$ is an effective measurement rate, $\epsilon$ is an
overall detection efficiency and $dW(t)$ a white noise Wiener increment. For the detailed expressions of $J$ and $\gamma$, cf. the main text or Supplementary Note~\ref{sec:continuous_readout} below. The corresponding homodyne current reads 
\begin{equation}
I(t)=2\sqrt{\epsilon\gamma}\langle\hat{H}/J\rangle_{c}+\xi(t),\label{eq:Ih_measureH}
\end{equation}
with $\xi(t)$  white (shot) noise  $dW(t)\equiv \xi(t)dt$.
We conclude Supplementary Note~\ref{sec:continuous_readout} with a brief discussion
on the filtering of the homodyne current.

In Supplementary Note~\ref{sec:experiment} we discuss some experimental considerations
on the proposed trapped-ion implementation, including the analysis
of its scalability, and the discussion of its robustness against major
experimental imperfections. Supplementary Note~\ref{sec:experiment} also provides
typical numbers for a proof-of-principle experiment.

We remark that our QND measurement scheme can be implemented with both
transverse ($x$ direction) and axial ($z$ direction) phonon modes of the 1D ion
string. While transverse phonon modes give rise to the power-law spin
interactions $J_{ij}\propto |i-j|^{-\alpha}$ with $1<\alpha<3$ as is considered
in the main text, axial phonon modes provide rich opportunities for engineering
exotic spin couplings~\cite{Porras2004}. The derivation of our scheme for both
cases is essentially the same. For notational concreteness, in the following
Supplementary Notes~\ref{subsec:Double-Molmer-Sorensen-interaction} and
\ref{sec:continuous_readout}, we derive the equations by assuming transverse
phonon modes. With the simple replacement $x\to z$ for the ionic motional
operators, the same derivation applies to the axial case. In
Supplementary Note~\ref{sec:experiment}, we discuss the features and experimental requirements
of the transverse and the axial implementation separately.

\subsection{Analytical study of the double M\o lmer-S\o rensen configuration}

\label{subsec:Double-Molmer-Sorensen-interaction} In this section we analyze in
detail the laser configuration which generates the desired system-meter coupling
$\hat{H}_{{\cal SM}}$ \eqref{eq:H_QND} as an effective Hamiltonian derived in
perturbation theory for the laser assisted spin-mode couplings.  While the model
Hamiltonian in the main text refers to the lowest order terms in this expansion,
we also derive the higher-order corrections to $\hat{H}_{{\cal SM}}$ and argue
that they are indeed negligible under typical experimental conditions.

\subsubsection{Light-ion coupling\label{subsec:Light-ion-coupling}}

We consider $N$ ions trapped in a linear Paul trap. The internal structure of
each ion is assumed to be a two level system (TLS), consisting of two qubit
states $\left|\downarrow\right\rangle $ and $\left|\uparrow\right\rangle $.  The
transition
$\left|\downarrow\right\rangle \rightarrow\left|\uparrow\right\rangle $ is
driven by two pairs of laser beams, such that each pair realizes a M\o lmer-S\o
rensen (MS) configuration, as shown schematically in Fig.~2 of the main
text. The lasers which form the first pair, shown as the amber beams, are
detuned by $\pm\Delta$ from the qubit transition frequency
$\omega_{\uparrow\downarrow}\equiv E_{\uparrow}-E_{\downarrow}$ respectively,
and have wave vector projections $\pm k$ along the $x$ direction; The lasers
corresponding to the second pair, shown as the blue beams, are detuned by
$\pm\Delta'$ from $\omega_{\uparrow\downarrow}$, and have wave vector projections
$\mp k$ along the $x$ direction. In the frame rotating at
$\omega_{\uparrow\downarrow}$, the full Hamiltonian of the internal and motional
degrees of freedom (DOFs) of the ion chain reads
\begin{equation}
\hat{H}_{\text{full}}=\hat{H}_{0}+\hat{V}.\label{eq:H_t}
\end{equation}
Here $\hat{H}_{0}$ is the Hamiltonian of the external motion of the ions (along
the $x$ direction), and can be expressed in terms of the collective phonon modes
\begin{equation}
\label{eq: phonon_free_H}
\hat{H}_{0}=\sum_{q}\omega_{q}\hat{a}_{q}^{\dagger}\hat{a}_{q}.
\end{equation}
where $\omega_{q}$ and $\hat{a}_{q}$ respectively denote the frequency and the
annihilation operator of the mode $q$. Here, the modes are ordered according to
their energy. For transverse phonon modes, the COM mode $q = 0$ has the highest
frequency, and therefore $\omega_{q} > \omega_{q+1}$. The order of modes is
reversed for axial phonon modes. The interaction between the ions and the lasers
is described by the Hamiltonian
\begin{align}
\hat{V}= & \frac{1}{2}\sum_{j=1}^{N}\hat{\sigma}_{j}^{+}\Big(\Omega_{1}e^{-i\Delta t+ik\hat{X}_{j}+i\zeta_{j}^{1}}+\Omega_{2}e^{i\Delta t-ik\hat{X}_{j}+i\zeta_{j}^{2}}\nonumber \\
 & +\Omega_{3}e^{-i\Delta't-ik\hat{X}_{j}+i\zeta_{j}^{3}}+\Omega_{4}e^{i\Delta't+ik\hat{X}_{j}+i\zeta_{j}^{4}}\Big)+{\rm {H.c}.}\label{eq:Hint_schro_pic}
\end{align}
Here $\Omega_{m}$ with $m = 1, \dotsc, 4$ denotes the Rabi frequency of the
laser beams, which is assumed to be real and positive for concreteness.  The
lasers with indices $m = 1,2$ correspond to the first MS pair and are shown as
amber beams in Fig.~2 of the main text. The blue beams, which correspond to the
lasers with indices $m = 3,4$, form the second MS pair. For the $j$-th ion,
$\hat{\sigma}_{j}^{+}\equiv\left|\uparrow\right\rangle _{j}\left\langle
  \downarrow\right|$
is its internal raising operator, and $\zeta_{j}^{m}$ is the phase of laser $m$
at its equilibrium position. The operator $\hat{X}_{j}$ describes the
small-amplitude displacement from the equilibrium position along the $x$
direction, and can be expressed in terms of the phonon operators as
$k\hat{X}_{j}=\sum_{q}\eta_{q}M_{jq}\left(\hat{a}_{q}+\hat{a}_{q}^{\dagger}\right)$,
where $M_{jq}$ is the distribution matrix element of mode $q$, and the
Lamb-Dicke (LD) parameters are defined as
$\eta_{q}=\eta\sqrt{\omega_{0}/\omega_{q}}$, with $\eta=k/\sqrt{2m\omega_{0}}$.

Hereafter we consider the Rabi frequencies of the four laser beams being
approximately equal up to a small offset,
\begin{align}
\Omega_{1} & =\Omega_{3}=\Omega,\nonumber \\
\Omega_{2} & =\Omega_{4}=\Omega+\delta\Omega.\label{eq:Rabi_frequency_choice}
\end{align}
According to Supplementary Note~\ref{subsec:Expansion-with-respect}, the small Rabi
frequency mismatch $\delta\Omega$ creates the desired transverse field term of
the Ising Hamiltonians~\eqref{eq:H_Ising_coupling_to_meter} and
\eqref{eq:H_Ising_intro}, with the transverse field strength
$h\propto\delta\Omega$. 

A pair of M\o lmer-S\o rensen laser beams is known to create the Ising spin
Hamiltonian Eq.~\eqref{eq:H_Ising_intro} in the off-resonant regime
$\Delta^{(\prime)}\gg\Omega$, $|\Delta^{(\prime)}-\omega_{q}|\gg\eta_{q}\Omega$
(see Refs.~\cite{Kim2009} and the discussion below). In our double MS
configuration, however, an additional term describing the QND coupling between
the Ising spin Hamiltonian and the COM phonon mode is generated {[}see the
second term of Eq.~\eqref{eq:H_QND}{]}. This is achieved by tuning
$\Delta^{\prime}=\Delta+\omega_{0}$, i.e., by choosing the beating between the
two pairs of MS lasers to match the COM phonon excitation frequency. It leads to
a resonant crosstalk between the two MS configurations, which results in the
desired QND coupling term.

In the following, we derive Eq.~\eqref{eq:H_QND} via a Magnus expansion of the
time evolution of the ion chain in the interaction picture. We shall first
introduce our method, which is a combined Magnus expansion and Lamb-Dicke
expansion, in Supplementary Notes~\ref{subsec:methodology_magnus}
and~\ref{subsec:Expansion-with-respect}, respectively, and then work out the
detailed expression of Eq.~\eqref{eq:H_QND} order by order. We summarize the
results in Supplementary Note~\ref{subsec:summary_and_tuning}.

\subsubsection{Magnus expansion: effective Hamiltonian}
\label{subsec:methodology_magnus}
Performing the gauge transformation $\hat{\sigma}_{j}^{+}\to\hat{\sigma}_{j}^{+}\exp\left[\left.-i\left(\zeta_{j}^{1}+\zeta_{j}^{2}\right)\right/2\right]$
and moving into the interaction picture with respect to $\hat{H}_{0}$,
Eq.~\eqref{eq:Hint_schro_pic} becomes 
\begin{align}
  \hat{V}_{I}= & \frac{\Omega}{2}\sum_{j=1}^{N}\hat{\sigma}_{j}^{+}\Big[ e^{-i\Delta t+ik\hat{X}_{j}\left(t\right)+i\varphi_{j}}\nonumber \\
               & +{\left( 1+\frac{\delta\Omega}{\Omega} \right)} e^{i\Delta t-ik\hat{X}_{j}\left(t\right)-i\varphi_{j}}\nonumber \\
               & +e^{-i\Delta't-ik\hat{X}_{j}\left(t\right)+i(\theta+\varphi_{j}^{\prime})}\nonumber \\
               & + \left( 1+\frac{\delta\Omega}{\Omega} \right) e^{i\Delta't+ik\hat{X}_{j}(t)+i(\theta-\varphi_{j}^{\prime})}\Big]+{\rm {H.c}.}\label{eq:H_int}
\end{align}
where the time-dependent position operator can be expressed in terms of the
phonon modes as
$k\hat{X}_{j}\left(t\right)=\sum_{q}\eta_{q}M_{jq}\hat{x}_{q}(t)$ with
$\hat{x}_{q}(t)\equiv\hat{a}_{q}{\rm exp}(-i\omega_{q}t)+{\rm H.c.}$, and the
relative laser phases are denoted as
$\theta=\left(\zeta_{j}^{3}+\zeta_{j}^{4}-\zeta_{j}^{1}-\zeta_{j}^{2}\right)/2,$
$\varphi_{j}=\left(\zeta_{j}^{1}-\zeta_{j}^{2}\right)/2,$ and
$\varphi_{j}'=\left(\zeta_{j}^{3}-\zeta_{j}^{4}\right)/2$.  We note that the
phase $\theta$ is independent of the ion index $j$. For the implementation with
transverse phonons considered here, this is simply because the laser phase
$\zeta_j^m$ is independent of $j$. For the axial implementation, this is also
true, as the position dependencies of the phases of two counter-propagating
lasers cancel each other.

We consider the regime where the MS lasers drive the qubit transition and the phonon sidebands
off-resonantly, $\Delta^{(\prime)}\gg\Omega$, $|\Delta^{(\prime)}-\omega_{q}|\gg\eta_{q}\Omega$.
The evolution operator corresponding to Eq.~\eqref{eq:H_int} can
be formally written as a Magnus series,
\begin{align}
\hat{U}(t) & \equiv\exp\left[-i\hat{G}(t)\right]=\text{\ensuremath{\mathcal{T}}exp}\left[-i\int_{0}^{t}dt_{1}\hat{V}_{I}\left(t_{1}\right)\right],\nonumber \\
\hat{G}\left(t\right) & =\sum_{l=1}^{\infty}\hat{G}_{l}(t).\label{eq:G}
\end{align}
Correspondingly, it allows us to define an effective Hamiltonian
$\hat{H}_{\text{eff}}\equiv\lim_{t\rightarrow\infty}\hat{G}\left(t\right)/t$
which describes the slow dynamics of the ion chain on a time scale much longer
than the phononic oscillation period $\sim 1/\omega_{0}$~\cite{Kim2009}.  The
lowest-order terms of the Magnus series are given by
\begin{align}
\hat{G}_{1}(t)  =&\int_{0}^{t}dt_{1}\hat{V}_{I}\left(t_{1}\right)dt_{1},\label{eq:G_1}\\
\hat{G}_{2}(t) =&-\frac{i}{2}\int_{0}^{t}dt_{1}\int_{0}^{t_{1}}dt_{2}\left[\hat{V}_{I}\left(t_{1}\right),\hat{V}_{I}\left(t_{2}\right)\right]\label{eq:G_2},\\
\hat{G}_{3}(t)  =&-\frac{1}{6}\int_{0}^{t}dt_{1}\int_{0}^{t_{1}}dt_{2}\int_0^{t_2}dt_3\nonumber\\
&\Big\{\left[\hat{V}_{I}\left(t_{1}\right),\left[\hat{V}_{I}\left(t_{2}\right),\hat{V}_I\left(t_3\right)\right]\right]\nonumber\\
&+
\left[\left[\hat{V}_{I}\left(t_{1}\right),\hat{V}_{I}\left(t_{2}\right)\right],\hat{V}_I\left(t_3\right)\right]
\Big\}.
\label{eq:G_3}
\end{align}
In the next section we derive $\hat{H}_{{\rm eff}}$ via explicit calculation of
$\hat{G}(t)$ in the long-time limit. In this calculation we further
perturbatively expand $\hat{V}_{I}$ in terms of the small Lamb-Dicke parameter
$\eta\ll1$. This allows us to construct $\hat{H}_{{\rm eff}}$ order by order as
a systematic expansion in $\eta$.

\subsubsection{Expansion with respect to the Lamb-Dicke parameter $\eta$}

\label{subsec:Expansion-with-respect} We now construct the effective
Hamiltonian $\hat{H}_{{\rm eff}}$ as an expansion with respect to
the Lamb-Dicke parameter, $\hat{H}_{\text{eff}}=\sum_{\ell=0}^{\infty}\hat{H}_{\text{eff}}^{(\ell)}$
with $\hat{H}_{\text{eff}}^{(\ell)}\propto\eta^{\ell}$. In
order to do that, we first expand the interaction Hamiltonian Eq.~\eqref{eq:H_int}
as $\hat{V}_{I}=\sum_{\ell=0}^{\infty}\hat{V}_{I}^{(\ell)}$, with
\begin{align}
\hat{V}_{I}^{(\ell)}\left(t\right)= & \frac{1}{2\times \ell !}\sum_{j=1}^{N}\hat{\sigma}_{j}^{+}\bigg[i\sum_{q}\eta_{q}M_{jq}\hat{x}_{q}\left(t\right)\bigg]^{\ell}\nonumber \\
 & \times\bigg[{\Omega}e^{-i\Delta t+i\varphi_{j}}+\left(-1\right)^{\ell}(\Omega+\delta\Omega)e^{i\Delta t-i\varphi_{j}}\nonumber \\
 & +\left(-1\right)^{\ell}\Omega e^{-i\Delta't+i(\theta+\varphi_{j}^{\prime})}\nonumber \\
 & +(\Omega+\delta\Omega)e^{i\Delta't+i(\theta-\varphi_{j}^{\prime})}\bigg]+{\rm {H.c}.}\label{eq:V_I_expansion}
\end{align}
To simplify the analysis, hereafter we consider $\theta=0$. Moreover, we choose
$\varphi_{j+1}-\varphi_{j}=2\pi s,$
$\varphi_{j+1}^{\prime}-\varphi_{j}^{\prime}=-2\pi s$ with
$s\in\mathbb{Z}$. For the implementation using transverse phonon modes, this
condition is automatically satisfied, with
$s=0$. For the implementation using axial phonon modes, this can be achieved by using the central part of an ion chain in a standard Paul trap
with nearly uniform spacing $d$ (or alternatively by using ions in
equal-distance ion traps~\cite{Schulz_2008,Harlander2011,Mehta2014,Wilson2014})
and by choosing an appropriate wavevector $k$ of the MS beams such that
$kd=2\pi s$. 

\subsubsection{Contributions from $\hat{G}_1(t)$}
Substituting the expression Eq.~\eqref{eq:V_I_expansion} into Eq.
\eqref{eq:G_2} and taking into account the conditions $|\Delta|\gg\Omega$ and
$|\Delta-\omega_{q}|\gg\eta_{q}\Omega$, we immediately see that $\hat{G}_{1}(t)$
does not contribute to $\hat{H}_{\text{eff}}$. Indeed, $\hat{G}_{1}(t)$
describes small-amplitude fast oscillations at frequency $\sim\Delta$, which
average to zero in the long-time regime.

\subsubsection{Contributions from $\hat{G}_2(t)$ up to $\eta^3$: $\hat{H}_{\cal SM}$}
\label{}
Below, we derive the contribution from $\hat{G}_{2}\left(t\right)$ as
an expansion in the Lamb-Dicke parameter. In this derivation, we implicitly
assume that the small offset of the Rabi frequency {[}see
Eq.~\eqref{eq:Rabi_frequency_choice}{]} satisfies
$\delta\Omega/\Omega\sim O(\eta_{q}^{2})$.

\paragraph{Transverse field terms.}

The zeroth-order expansion of $\hat{H}_{{\rm eff}}$ is readily constructed by
plugging $\hat{V}_{I}^{(0)}$ into Eq.~\eqref{eq:G_2},
\begin{align}
\hat{H}_{\text{eff}}^{\left(0\right)} & =-\frac{i}{2t}\int_{0}^{t}dt_{1}\int_{0}^{t_{1}}dt_{2}\left[\hat{V}_{I}^{\left(0\right)}\left(t_{1}\right),\hat{V}_{I}^{\left(0\right)}\left(t_{2}\right)\right]\nonumber \\
 & =h\sum_{j=1}^{N}\hat{\sigma}_{j}^{z}.\label{eq:H^0}
\end{align}
We note that here and below, we implicitly assume the long-time limit
$t \to \infty$. The zeroth-order contribution provides the transverse field term
of the quantum Ising Hamiltonian, with the transverse field strength given by
\begin{equation}
h=\frac{\Omega\delta\Omega}{2}\left(\frac{1}{\Delta}+\frac{1}{\Delta^{\prime}}\right).\label{eq:transverse_field}
\end{equation}
Similarly, the first-order expansion $\hat{H}_{{\rm eff}}^{(1)}$ is given by
\begin{align}
\hat{H}_{\text{eff}}^{\left(1\right)} & =-\frac{i}{2t}\sum_{\ell+m=1}\int_{0}^{t}dt_{1}\int_{0}^{t_{1}}dt_{2}\left[\hat{V}_{I}^{(\ell)}\left(t_{1}\right),\hat{V}_{I}^{(m)}\left(t_{2}\right)\right]\nonumber \\
 & =-\vartheta h\sum_{j=1}^{N}\hat{\sigma}_{j}^{z}\otimes\hat{P},\label{eq:H^1}
\end{align}
where $\hat{P}\equiv i(\hat{a}_{0}e^{i\varphi}-\hat{a}_{0}^{\dagger}e^{-i\varphi})/\sqrt{2}$
is a quadrature operator of the COM phonon mode, with $\varphi\equiv\varphi_{j}-\varphi_{j}'$
an angle dependent on the laser phases, and $\vartheta=-\eta_{0}\sqrt{2}M_{i0}\simeq-\eta_{0}\sqrt{2/N}$
the dimensionless coupling strength. 
In the following, we absorb the phase $\varphi$ into the definition
of $\hat{a}_{0}$, $\hat{a}_{0}e^{i\varphi}\to-\hat{a}_{0}$, thus
$\hat{P}=i(\hat{a}_{0}^{\dag}-\hat{a}_{0})/\sqrt{2}$.

\paragraph{Ising terms.}

The second order expansion of $\hat{H}_{{\rm eff}}$ can be constructed
analogously,
\begin{align}
\hat{H}_{\text{eff}}^{\left(2\right)} & =-\frac{i}{2t}\sum_{\ell+m=2}\int_{0}^{t}dt_{1}\int_{0}^{t_{1}}dt_{2}\left[\hat{V}_{I}^{(\ell)}\left(t_{1}\right),\hat{V}_{I}^{(m)}\left(t_{2}\right)\right]\nonumber \\
 & =-\sum_{i<j}J_{ij}\hat{\sigma}_{i}^{x}\hat{\sigma}_{j}^{x}.\label{eq:H^2}
\end{align}
In this derivation we dropped terms $\sim\eta^{2}\delta\Omega$ under our
assumption $\delta\Omega/\Omega\propto\eta^{2}$. We discuss the effect of these
higher-order terms in Supplementary Note~\ref{sec:higher-order_correction}.
Equation~\eqref{eq:H^2} describes an Ising spin-spin coupling with coupling
strength
\begin{align}
J_{ij}= & -\Omega^{2}\sum_{q}\eta_{q}^{2}\omega_{q}M_{iq}M_{jq}\nonumber \\
 & \times\left[\frac{1}{\Delta^{2}-\left(\omega_{q}\right)^{2}}+\frac{1}{(\Delta')^{2}-\left(\omega_{q}\right)^{2}}\right],\label{eq:J_mn}
\end{align}
which includes two independent contributions from the two MS laser
configuration. 

Finally, the third order expansion of $\hat{H}_{{\rm eff}}$ can be calculated in
an analogous (though lengthy) way
\begin{align}
\hat{H}_{\text{eff}}^{\left(3\right)} & =-\frac{i}{2t}\sum_{\ell+m=3}\int_{0}^{t}dt_{1}\int_{0}^{t_{1}}dt_{2}\left[\hat{V}_{I}^{(\ell)}\left(t_{1}\right),\hat{V}_{I}^{(m)}\left(t_{2}\right)\right]\nonumber \\
 & =-\vartheta\Bigg(\sum_{i<j}J_{ij}\hat{\sigma}_{i}^{x}\hat{\sigma}_{j}^{x}+\mathcal{E}\Bigg)\otimes\hat{P}.\label{eq:H^3}
\end{align}
Here, the spin-spin interaction strength $J_{ij}$ is defined in Eq.~\eqref{eq:J_mn},
and $\vartheta$ and $\hat{P}$ are defined below Eq.~\eqref{eq:H^1}.
$\mathcal{E}\equiv\sum_{j}J_{jj}/2$ is a constant driving field for
the COM quadrature, which we neglect in the following as it just
leads to a constant component in the measured signal. Equation~\eqref{eq:H^3}
results from a resonant cross-talk between the two MS laser configuration
under the condition $\Delta'=\Delta+\omega_{z}$, and describes the
QND coupling between the spin Hamiltonian and the quadrature of the
COM phonon mode.

In deriving Eq.~\eqref{eq:H^3}, an important assumption we made
is that no other phonon mode is resonantly excited except the COM mode up to third order
in the Lamb-Dicke parameter. This ``single sideband addressibility'' is guaranteed by the condition 
\begin{equation}
\sum_q \frac{\eta^2_q\Omega^2\omega_q\eta_p}{[\Delta^{(\prime)}]^2-\omega_q^2}\ll |\omega_p-\omega_0|,\quad \quad\forall p\neq 0.
\label{eq:validity}
\end{equation}
Equation~\eqref{eq:validity} can be interpreted physically as follows:
In our scheme, the excitation of sideband phonons is achieved by
simultaneously flipping \emph{two} spins. The strength for simutaneously flipping spin $i$ and $j$ is given by $J_{ij}$ in Eq.~\eqref{eq:J_mn}. As a result, the sideband addressing strength is $\sim \eta \Omega^2\sum_q \eta^2_q\omega_q/\{[\Delta^{(\prime)}]^2-\omega_q^2\}$ (we set $M_{iq}=1$ for a worst-scenario analysis), and should be much smaller than the spectral gap between the COM mode and other modes.

%

In a transverse-phonon implementation, the phonon spectrum gets denser for increasing number of ions. Thus the validity of condition~\eqref{eq:validity}
sets a limit on the scalability of our QND scheme. This is analyzed in detail later in Supplementary Note~\ref{sec:experiment}. Here, we only note that Eq. \eqref{eq:validity} is much less stringent than the requirement for sideband addressing via laser flipping \emph{individual} spins, $\eta_q \Omega \ll |\omega_q-\omega_0|,\quad\forall q\neq 0$, as is required, e.g., by the Cirac-Zoller gate or the near-resonant M\o lmer-S\o rensen gate. This leads to nice scalability of our QND measurement scheme for a given laser power.

Combining Eqs. \eqref{eq:H^0}, \eqref{eq:H^1}, \eqref{eq:H^2} and
\eqref{eq:H^3}, the effective Hamiltonian of the ion chain $\hat{H}_{{\rm eff}}$
can be written in the form of $\hat{H}_{\cal SM}$ in Eq.~\eqref{eq:H_QND}, with the identification
\begin{align}
\hat{H} & =-\sum_{i<j}J_{ij}\hat{\sigma}_{i}^{x}\hat{\sigma}_{j}^{x}-h\sum_{j=1}^{N}\hat{\sigma}_{j}^{z},\nonumber \\
\hat{H}' & =-\text{\ensuremath{\sum_{i<j}J_{ij}\hat{\sigma}_{i}^{x}\hat{\sigma}_{j}^{x}}}+h\sum_{j=1}^{N}\hat{\sigma}_{j}^{z},\label{eq:spinHamiltonian_details}
\end{align}
$\vartheta\simeq-\eta_{0}\sqrt{2/N}$ and $\hat{P}=i(\hat{a}_{0}^{\dag}-\hat{a}_{0})/\sqrt{2}$. 

\subsubsection{Tuning of $\hat{H}_{{\cal SM}}$}
\label{subsec:summary_and_tuning}
Here we describe a method to further tune the transverse field in $\hat{H}$ and
$\hat{H}'$ {[}cf. Eq.~\eqref{eq:spinHamiltonian_details}{]} independently, thus
allowing to reach the QND sweetspot $\hat{H}=\hat{H}'$.  To this end, we
consider the same laser configuration as in
Supplementary Note \ref{subsec:Light-ion-coupling}, nevertheless the detunings of the MS
lasers are now respectively modified to $B\pm\Delta,$ $B\pm\Delta^{\prime}$,
with $B\sim J\ll\Delta,\Delta^{\prime}$.  In the frame rotating at frequency
$\omega_{\uparrow\downarrow}+B$, we get an additional term
$B\sum_{j=1}^{N}\hat{\sigma}_{j}^{z}$ in the Hamiltonian of the laser-driven ion
chain $\hat{H}_{\text{full}}$ {[}cf. Eq.~\eqref{eq:H_t}{]}. Repeating the same
derivation as described in Supplementary Notes \ref{subsec:Light-ion-coupling} and
\ref{subsec:Expansion-with-respect}, we recover exactly the same $\hat{H}$ that
is coupled to the meter DOFs, while $H_{}^{\prime}$ is modified as
\[
H^{\prime}=\text{\ensuremath{\sum_{i<j}J_{ij}\hat{\sigma}_{i}^{x}\hat{\sigma}_{j}^{x}}}-\left(h-B\right)\sum_{j}\hat{\sigma}_{j}^{z}.
\]
By choosing $B=2h$ we realize the QND condition $\hat{H}^{\prime}=\hat{H}$,
while offsetting $B$ slightly from $2h$ allows us to observe quantum
jumps between different eigenstates of $\hat{H}$.

\subsubsection{Contributions from $\hat{G}_2(t)$ beyond $\eta^3$: higher-order corrections}

\label{sec:higher-order_correction} In this section we derive the corrections to
the QND Hamiltonian Eq.~\eqref{eq:H_QND} resulting from higher-order
terms in the Lamb-Dicke expansion of $\hat{G}_2(t)$. We show that these terms do
not change the QND character of the proposed measurement scheme.

By straight forward calculation, we find that the fourth order expansion of the
effective Hamiltonian can be written as
\begin{align}
\hat{H}_{\text{eff}}^{\left(4\right)} & =-\frac{i}{2t}\sum_{\ell+m=4}\int_{0}^{t}dt_{1}\int_{0}^{t_{1}}dt_{2}\left[\hat{V}_{I}^{(\ell)}\left(t_{1}\right),\hat{V}_{I}^{(m)}\left(t_{2}\right)\right]\nonumber \\
 & =-\sum_{i<j}J_{ij}^{(4)}\hat{\sigma}_{i}^{x}\hat{\sigma}_{j}^{x},\label{eq:H^4}
\end{align}
with the spin-spin coupling 
\begin{align}
J_{ij}^{(4)}= & -\frac{\Omega^{2}}{2}\sum_{qp}\eta_{q}^{2}\eta_{p}^{2}M_{iq}M_{jq}M_{ip}M_{jp}(\omega_{q}+\omega_{p})\nonumber\\
 & \times\left[\frac{1}{\Delta^{2}-(\omega_{q}+\omega_{p})^{2}}+\frac{1}{(\Delta')^{2}-(\omega_{q}+\omega_{p})^{2}}\right]\nonumber\\
 & +\frac{\Omega^{2}}{2}\eta_{0}^{2}\omega_{z}\sum_{p}\eta_{p}^{2}(M_{ip}^{2}+M_{jp}^{2})\nonumber\\
 & \times\sum_{q}M_{iq}M_{jq}\left[\frac{1}{\Delta^{2}-\omega_{q}^{2}}+\frac{1}{(\Delta')^{2}-\omega_{q}^{2}}\right].\label{eq:Jij4}
\end{align}
In the derivation of Eq.~\eqref{eq:H^4}, we dropped terms proportional
to the phonon-occupation under the assumption $\langle\hat{a}_{q}^{\dag}\hat{a}_{q}\rangle\ll1$.

Besides $\hat{H}_{\text{eff}}^{\left(4\right)}$, another correction
to the QND Hamiltonian that is fourth order in $\eta$ comes from
the term $\sim\eta^{2}\delta\Omega\propto\eta^{4}$ which
we dropped in Eq.~\eqref{eq:H^2}. Via straightforward calculation
we find this term can be written as a transverse field Ising Hamiltonian
with site-dependent transverse field, 
\begin{equation}
-\sum_{i<j}t_{ij}\hat{\sigma}_{i}^{x}\hat{\sigma}_{j}^{x}-\sum_{j}\lambda_{j}\hat{\sigma}_{j}^{z},\label{eq:Hcorrection}
\end{equation}
with the coefficients 
\begin{align}
t_{ij}= & -\Omega\delta\Omega\sum_{q}\omega_{q}\eta_{q}^{2}M_{iq}M_{jq}\nonumber \\
 & \times\left[\frac{1}{\Delta^{2}-\omega_{q}^{2}}+\frac{1}{(\Delta')^{2}-\omega_{q}^{2}}\right],\nonumber \\
\lambda_{j}= & -2\Omega\delta\Omega\sum_{q}\eta_{q}^{2}M_{jq}^{2}\left(\frac{1}{\Delta}+\frac{1}{\Delta'}\right).
\end{align}

Importantly, the corrections to the QND Hamiltonian up to fourth order
in the Lamb-Dicke parameter, Eq.~\eqref{eq:H^4} and~\eqref{eq:Hcorrection},
only involve spin DOFs and do not involve phonon DOFs. Thus, they
only slightly renormalize the coefficients of $\hat{H}'$ {[}cf. Eq.~\eqref{eq:spinHamiltonian_details}{]},
introducing tiny mismatch between $\hat{H}'$ and $\hat{H}$. As described
in the main text, these mismatch only introduce rare quantum jumps
between energy eigenstates {[}cf. Fig.~1(3) of the main text{]},
whereas the QND character of the measurement is maintained.

\subsubsection{Effects of higher order Magnus series}

In this section, we show that the contributions to the effective Hamiltonian from
higher-order terms in the Magnus series, $\hat{G}_{n}(t),n\geq 3$, are of higher order than $\eta^4$. Thus they are much smaller than the system-meter coupling
$\hat{H}_{\cal SM}$ and do not change the QND character of the
proposed measurement scheme. The following analysis is based on power counting
and physical arguments.

Let us first consider the contributions from $\hat{G}_3(t)$,
cf.~Eq.~\eqref{eq:G_3}. For simplicity, we temporarily assume balanced MS
configurations, i.e., $\delta\Omega=0$. In this case, the spin operators
$\hat{\sigma}_{j}^\pm$ in Eq.~\eqref{eq:V_I_expansion} can be combined to
$\hat{\sigma}_j^x$. Therefore, spin operators always commute with each other at
different time. As a result, the double commutator in the integrand of
Eq.~\eqref{eq:G_3} is nonzero only if both the inner and outer commutators
contain at least a pair of phonon annihilation and creation operators. Such an
integrand, contains at least four phonon operators, and its contribution is
$O(\eta^4)$ after the integration. Moreover, it is straightforward to see such a contribution contains a fast oscillating phase $\sim \Delta^{(\prime)} t$ and do not contribute to the effective Hamiltonian. Physically, it corresponds to processes which
involve two virtual phonon excitations, which are nevertheless off-resonant. Thus, $\hat{G}_3(t)$ contributes $O(\eta^5)$ to the effective Hamiltonian when $\delta\Omega=0$.

We now reintroduce the imbalance $\delta\Omega$. Keeping in mind that
$\delta\Omega/\Omega\sim O(\eta^2)$, in the integrand of
Eq.~\eqref{eq:V_I_expansion} we can restrict ourselves to `relevant' terms that contain $\delta\Omega$ and at most two phonon operators, since the other terms contribute at
$O(\eta^5)$ after multiplication with $\delta\Omega$. Under the conditions
$\Delta^{(\prime)}\gg\Omega$ and
$|\Delta^{(\prime)}-\omega_{q}|\gg\eta_{q}\Omega$, which are met in our off-resonant double MS configuration, it is straightforward to verify that the `relevant' terms involving zero or one phonon operator contain fast oscillating phase $\sim \omega_z t$ and average to zero in the long time limit. The analysis of the terms consisting of $\delta\Omega$ and two phonon operators is more involved. First, we note that these terms are of the order $O(\eta^4)$, i.e., much smaller than the QND coupling Hamiltonian $\hat{H}_{\cal SM}$. Secondly, we find that they are off-resonant and average to zero if we avoid `accidental' resonances, by requiring $|\omega_q-\omega_p|\neq |\Delta-\omega_0|,\, \forall p,q$, which can be achieved by choosing appropriate $\Delta$ in experiments. To summarize, the contribution of $\hat{G}_3(t)$ can be made as small as $O(\eta^5)$ and is thus negligible.

Next, we consider the contributions from $\hat{G}_4(t)$. Adopting similar
arguments as above, we find that if $\delta\Omega=0$ the contribution from
$\hat{G}_4(t)$ is on the order of $O(\eta^6)$, and corresponds physically to
processes involving three virtual phonon excitations. Reintroducing
$\delta\Omega$ and looking at `relevant' terms that contain at most two phonon
operators, we find only one term that does not oscillate and remains finite in
the long-time limit, under the assumption that accidental resonances are avoided and the phonon occupations are small, $\langle\hat{a}_{q}^{\dag}\hat{a}_{q}\rangle\ll1$. This term is proportional to
$\delta \Omega\times\Omega^3/\Delta^3\sum_j \hat{\sigma}_j^z$ and describes the
AC-Stark shift from fourth-order perturbation theory. Comparing this term to the
QND coupling terms \eqref{eq:H^1} and \eqref{eq:H^3}, we find it has a relative
strength $\Omega^2/(\eta\Delta^2)$. To ensure that this term has negligible
effect on the performance of our QND scheme, we thus require
$\Omega^2/(\eta\Delta^2)\ll 1$. This is typically true, as $\eta$ and
$\Omega/\Delta$ are small parameters which have similar magnitudes.

%

Along the same lines, we can show that the contributions from $\hat{G}_n,n> 4$
are all negligible. 

\subsection{Numerical study of the double M\o lmer-S\o rensen configuration}\label{sec: Numerical}

We complement the analytical investigation of the double M\o lmer-S\o rensen
laser configuration scheme above by a numerical study of the evolution of a
system of $N=3$ ions in a truncated phonon basis, and we identify parameter
regimes in which a description of the system in terms of the effective
Hamiltonian Eq.~\eqref{eq:H_QND} is valid. We consider two scenarios: First, we
study the evolution with $\delta \Omega = 0$, which corresponds to the Ising
model with no transverse field $h=0$, and we assume that the QND scheme is
implemented with transverse phonon modes. Under these conditions, we show that
in a wide range of detunings $\Delta$ the exact dynamics of the joint system
consisting of spin and phonon degrees of freedom modes is well reproduced by the
effective Hamiltonian Eq.~\eqref{eq:H_QND}. Second, for a fixed value of the
detuning $\Delta$, we show that the effective Hamiltonian remains valid for a
range of values of the transverse field $h$. This analysis assumes the use of
axial phonon modes and is based on Floquet theory.

\subsubsection{Dynamics of phonon modes}
The starting point of our consideration is the full Hamiltonian of the system Eq. \eqref{eq:H_int}. We focus on the transverse phonon modes and analyse the tunability of the detuning $\Delta$ which defines the spin-spin interaction parameter $\alpha$ as discussed.
We assume all Rabi frequencies to be equal ($ \delta \Omega=0$), which corresponds to the Ising model with no transverse field  Eq.~\eqref{eq:transverse_field}. In this case the full Hamiltonian simplifies to:
\begin{equation}
\hat{V}_I=\sum_{j=1}^{N} \sigma^{x}_{j} \Big[ \cos \Big(\Delta t-k \hat{X}_{j} (t) \Big)  + \cos \Big(\Delta' t+k \hat{X}_{j}(t) \Big) \Big]. \label{eq:ReducedH}
\end{equation}
This implies that for an initial product state in the $\sigma^x$ 
basis there is no dynamics of spin variables, and only the phonon modes evolve in time. 
If they are initially prepared in their vacuum states then, according to the effective 
Hamiltonian given in Eqs.~\eqref{eq:H_QND} and~\eqref{eq:H_Ising_coupling_to_meter}, at some final time 
$t_f$ we expect $\langle\hat{x}_q\rangle \propto \sqrt{\langle a_q^{\dag} a_q \rangle} \propto \delta_{q,0}(\sum J_{ij} s^x_i s^x_j )\cdot t_f $, where $\{s^x_i\}$ 
denote the initial spin configuration. 

To check the validity of our model we compute the mean phonon number
$ \langle a_q^{\dag} a_q \rangle$ using the full Hamiltonian
Eq.~\eqref{eq:ReducedH} for the corresponding initial spin states as a function
of the detuning $\Delta$, while keeping $\Delta'=\Delta+\omega_0$. The result is
shown in Supplementary Fig.~\ref{Fig:verification0}(a).  We observe the appearance of
resonances at certain values of $\Delta$, where the evolution does not
correspond to the effective Hamiltonian Eq.~\eqref{eq:H_QND}.  Partially this
can be explained by the poles of the fourth order expansion term of the $J_{ij}$
matrix Eq.~\eqref{eq:Jij4}.  The rest of resonances have lower amplitude and
higher frequency and we attribute them to the many-phonon resonant processes
which correspond to the higher-than-fourth order $\eta$-expansion terms.
 
Remarkably, there exist wide regions of detuning $\Delta$ free of resonances
provided that there is space between the higher order harmonics of the phononic
eigenfrequencies $\omega_0\gg |\omega_0-\omega_{q}|,\,\forall q$. As shown in
Supplementary Fig.~\ref{Fig:verification0}(b) the system dynamics in these regions is well
reproduced by the effective model Eqs.~\eqref{eq:H_QND},~\eqref{eq:J_mn} such
that the COM mode population reflects the eigenenergies of the spin system.

In order to show that the scheme scales well with the number of ions we present in Supplementary Fig.~\ref{Fig:verification0_6}(a,b) results for $N=6$ ions interacting via 6 phonon modes. One can see that effective QND dynamics represents the exact results as expected.
 
The setup which is described in this section is also of interest as a first
proof-of-principle experimental test of the proposed QND coupling of the spin
model to the COM phonons $\vartheta\hat{H}\otimes\hat{P}$. Importantly, it does
not require implementation of the full QND scheme with the continuous readout as
the mean phonon-number measurement can be done in a multi-shot fashion.

\begin{figure}
\includegraphics[width=1\columnwidth]{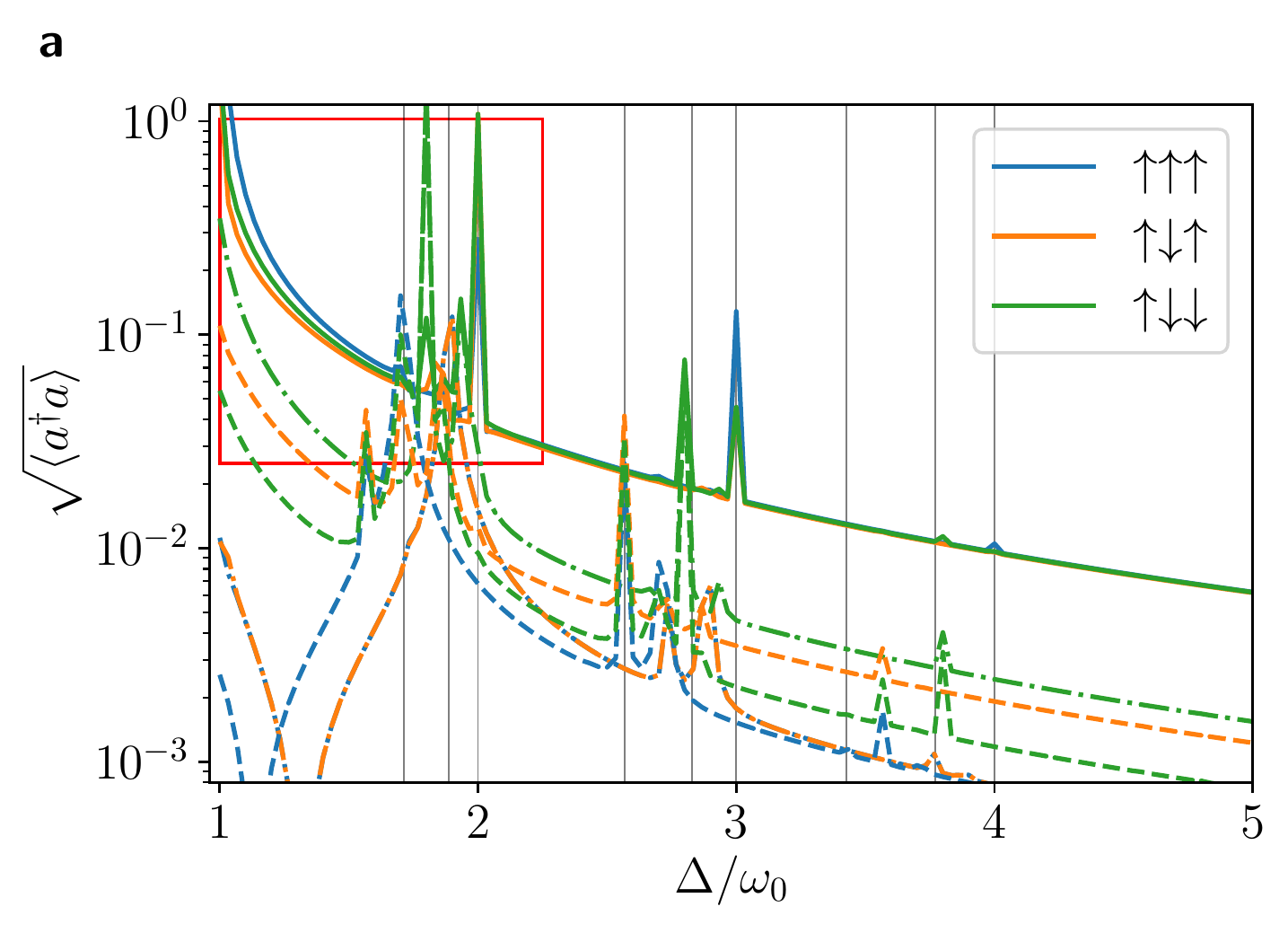}
\includegraphics[width=1\columnwidth]{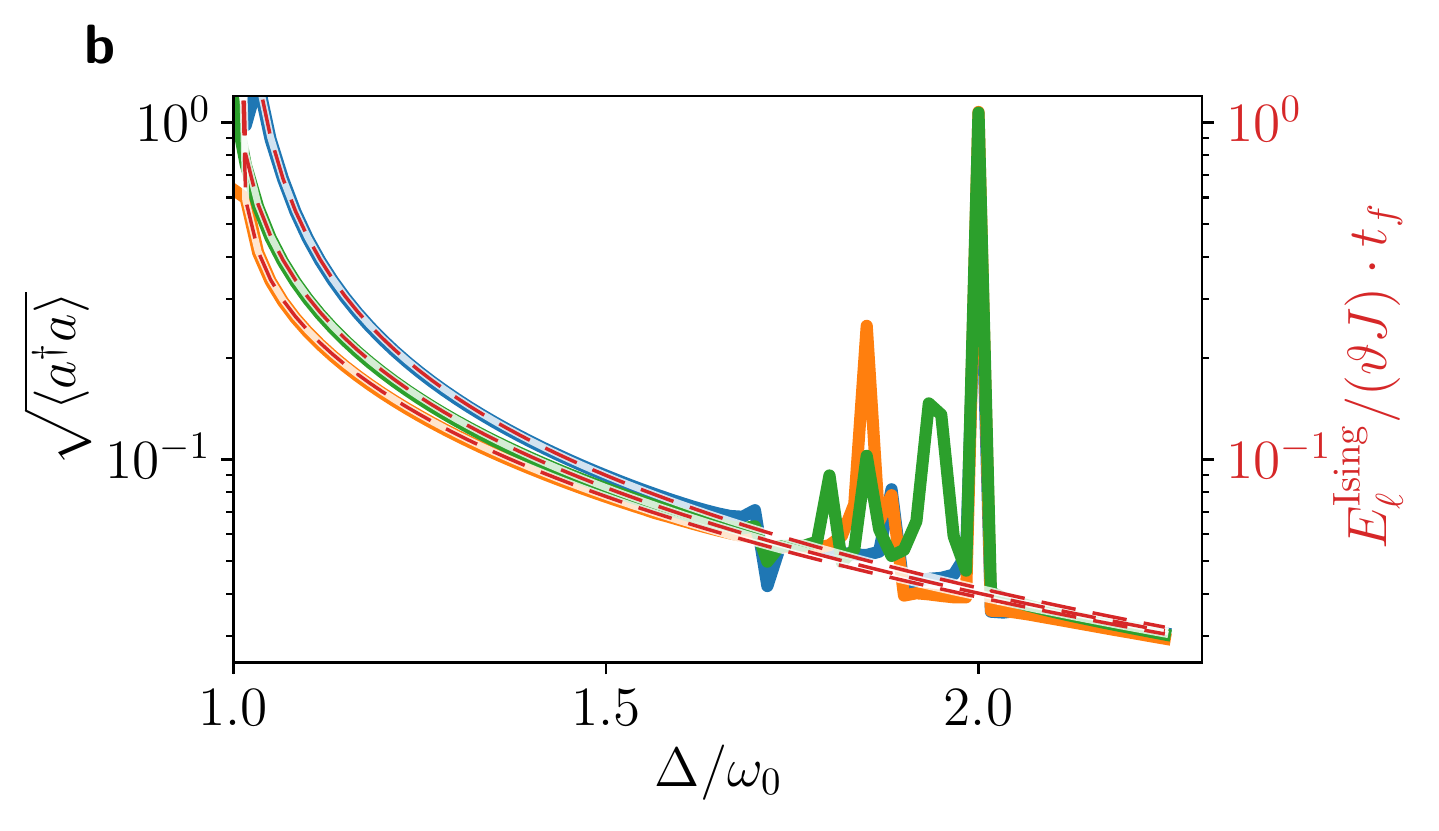}
\caption{(a) Transverse phonon mode occupation at the final time
  $t_f=2\pi \times 600/\omega_0$ for different initial states of $N=3$ spins indicated
  with blue, green, and orange colors. Solid, dot-dashed and dashed curves stand
  for the $q=0,1,2$ modes, respectively. Vertical lines represent
  frequencies of the higher-order harmonics of the phonon modes.  (b) Zoomed-in
  part [red rectangle in (a)] including the first resonances.  Dashed curves
  show the analytical results obtained from
  Eqs.~\eqref{eq:H_QND},~\eqref{eq:J_mn}. The numerical parameters:
  $\eta=0.1$, $\Omega/\omega_0=1/6,  \omega_0/\omega_z=3$, the phonon modes are described by 5 Fock
  states.  } 
\label{Fig:verification0}
\end{figure}

\begin{figure}
\raggedleft
\includegraphics[width=1\columnwidth]{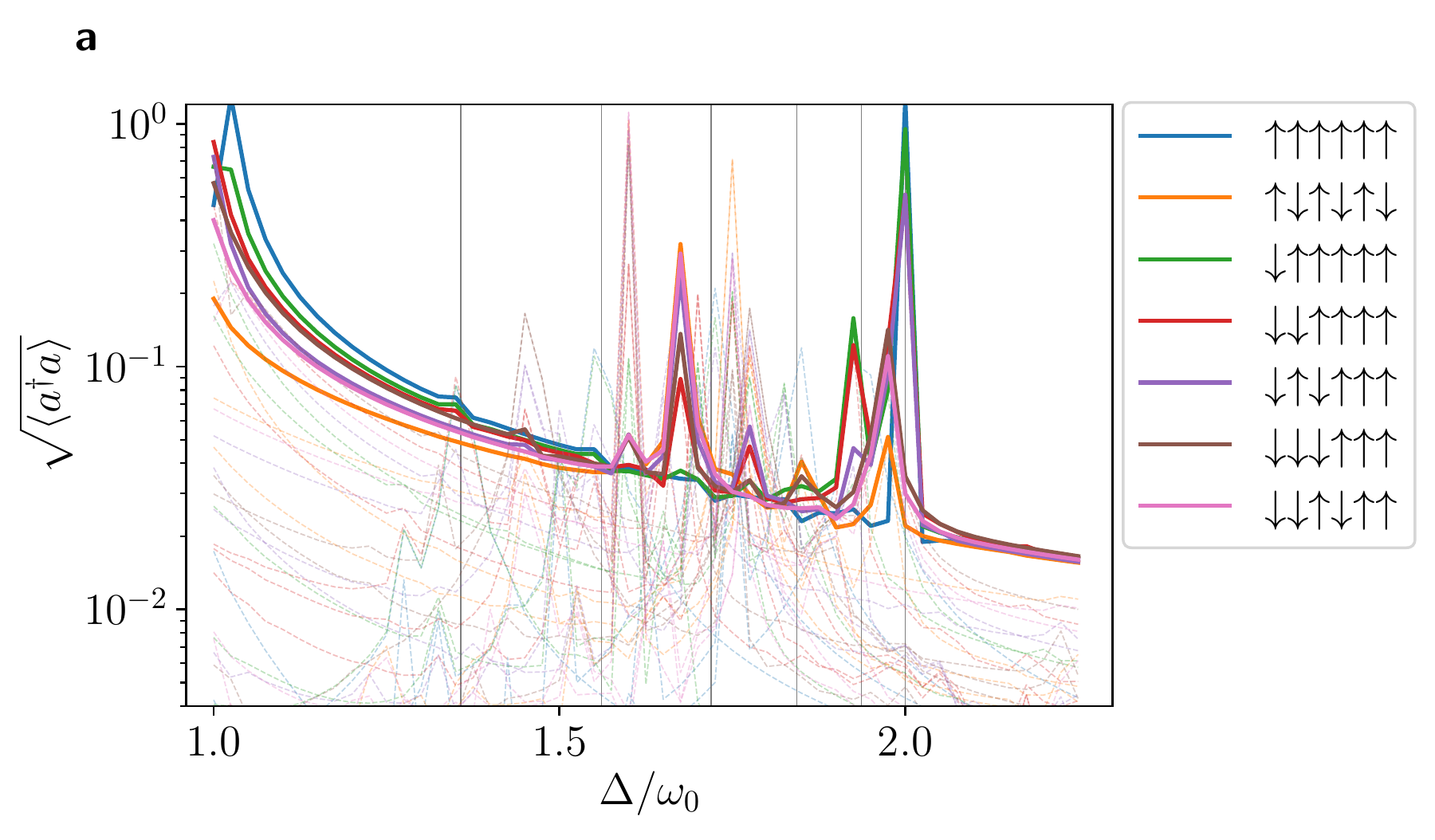}
\includegraphics[width=1\columnwidth]{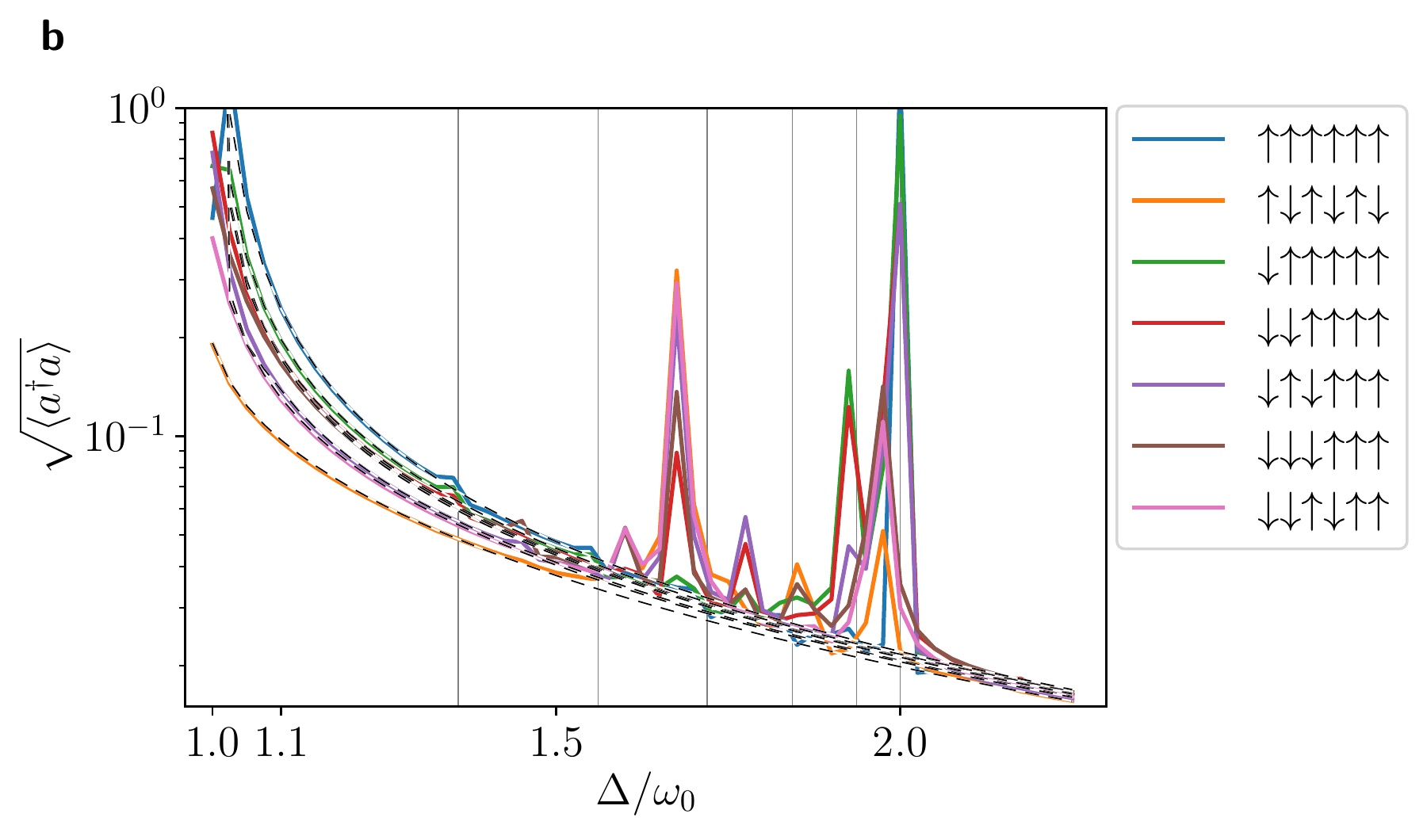}
\caption{(a) Transverse phonon mode occupation at the final time
  $t_f=2\pi \times 400/\omega_0$ for different initial states of $N=6$ spins indicated
  with various colors. Solid curves stand
  for the COM mode ($q=0$) population, dashed curves represent the rest of the modes $q\neq0$. Vertical lines represent
  frequencies of the higher-order harmonics of the phonon modes.  (b) Comparison with the analytical results.  Dashed curves
  show the analytical results obtained from
  Eqs.~\eqref{eq:H_QND},~\eqref{eq:J_mn}. Parameters:
  $\eta=0.1$, $\Omega/\omega_0=1/8,  \omega_0/\omega_z=4$. The phonon modes are described by 3 Fock
  states. }
\label{Fig:verification0_6}
\end{figure}

\subsubsection{Floquet spectrum analysis}

Here we verify the effective QND dynamics for various transverse field
values. In particular, we perform a numerical simulation of the periodically
driven system of $N=3$ ions interacting via 3 axial phonon modes according to
the full Hamiltonian $\hat{H}_{{\rm full}}(t)$ given by Eq.~\eqref{eq:H_t}.  We
choose commensurate detunings $\Delta=-7\omega_{0}$, $\Delta'=-6\omega_{0}$,
such that the overall dynamics is periodic with frequency $\omega_{0}$.  Next,
the operator of unitary evolution
$\hat{U}(t)=\mathcal{T}\exp\left[-i\int_{0}^{t}dt_{1}\hat{H}_{{\rm
      full}}(t_{1})\right]$
is numerically evaluated for one period of the oscillation. The logarithm of
eigenvalues of $\hat{U}(2\pi/\omega_{0})$ provides $E_{\ell}^{{\rm Floquet}}$
the quasi energies of the effective Hamiltonian.

In Supplementary Fig.~\ref{Fig:verification}(a) we compare the Floquet quasi energies
$E_{\ell}^{{\rm Floquet}}$ (blue dotted lines) with the spectrum
$E_{\ell}^{{\rm Ising}}$ of the effective Ising
Hamiltonian~\eqref{eq:H_Ising_intro} with adjusted transverse field $B=2h$
(dashed lines) for various values of the Rabi frequency mismatch $\delta\Omega$
expressed as a transverse field $h$ via Eq.~\eqref{eq:transverse_field}. The
figure clearly shows that the exact eigenvalues $E_{\ell}^{{\rm Floquet}}$ are
well represented by the effective Ising model.

Next, we study the coupling of the Ising Hamiltonian to the COM phonon
mode. Here we consider the non-hermitian Hamiltonian of the full system
(ions+phonons) $\hat{H}_{{\rm full}}(t)-i\frac{\gamma_{s}}{2}a_{0}^{\dagger}a_{0}$
with the non-hermitian term describing the decay of the COM mode due
to the read-out. The Floquet eigenstates with the quasi energies around
0 and small imaginary parts represent the steady states of the open
system. The COM mode displacements $\braket{a_{0}+a_{0}^{\dagger}}$
averaged over these Floquet states are shown in Supplementary Fig.~\ref{Fig:verification}(b)
with red lines. The displacement is proportional to the corresponding
eigenenergy of the Ising Hamiltonian~\eqref{eq:H_Ising_coupling_to_meter}
shown with dashed lines. The resulting read-out photocurrent is sensitive
to the amplitude of the COM mode oscillations and, therefore, reveals
the eigenenergies of the desired Ising model.

\begin{figure}
  \includegraphics[width=1\columnwidth]{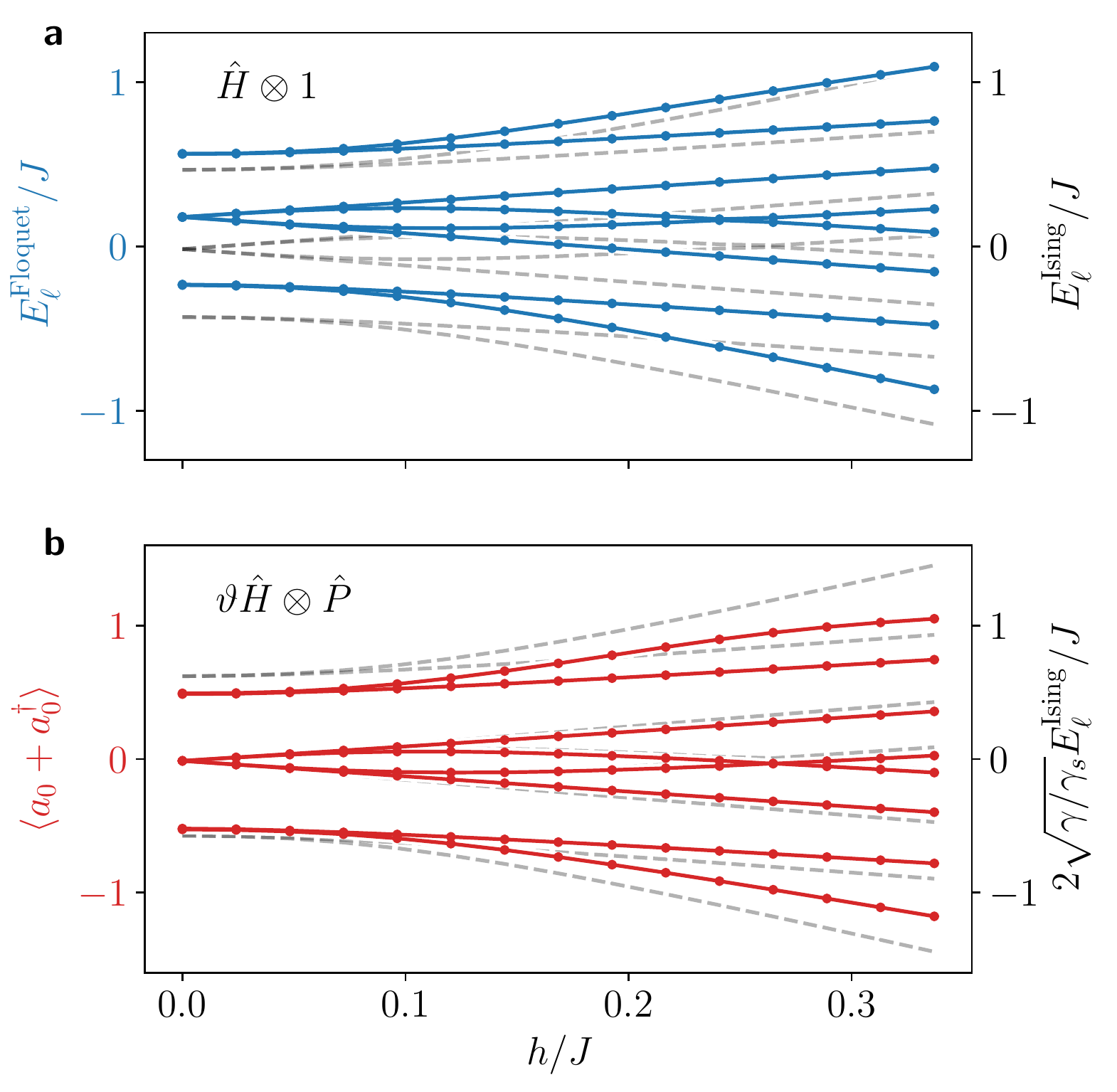}\caption{Numerical
    verification of the double M\o lmer-S\o rensen configuration via Floquet
    analysis. (a) Test of the free evolution term $\hat{H}\otimes\mathbb{{I}}$
    of the effective system-meter Hamiltonian~\eqref{eq:H_QND}. Blue lines show
    the exact Floquet spectrum depending on the effective transverse field $h$
    (see text), dashed lines represent eigenenergies of the effective Ising
    model. (b) The system-meter coupling $\vartheta\hat{H}\otimes\hat{P}$
    test. Red lines show the COM mode displacement
    $\braket{a_{0}+a_{0}^{\dagger}}$ averaged over the exact Floquet eigenstates
    (see text), dashed lines represent the corresponding eigenenergies of the
    effective Ising model.  The following parameters are used: $\eta=0.3$,
    $\Delta=-7\omega_{0}$, $\Delta'=-6\omega_{0}$, $\Omega=\omega_{0}$,
    $0<\frac{\delta\Omega}{\Omega}<2\times10^{-3}$,
    $\gamma_{s}=3\frac{\eta
      J}{\sqrt{N}}\left(\frac{\eta\Omega}{\Delta}\right)^{2}$.
    The COM mode is described by 6 Fock states, the other modes use 3 Fock
    states.}
\label{Fig:verification}
\end{figure}

\subsection{Continuous readout of the spin Hamiltonian}

\label{sec:continuous_readout} With the implementation of the system-meter
coupling Hamiltonian Eq.~\eqref{eq:H_QND} at hand, in this section
we present the detailed discussion on the readout of the transverse
Ising Hamiltonian via continuous monitoring the center-of-mass phonon
quadrature $\hat{X}$, extending the short description presented in
the Method section of the main text.

The experimental setup we have in mind is shown schematically in Fig.~2
of the main text. Here, aside from the ions $j\in\{1, \dotsc, N\}$ which generates
the QND Hamiltonian Eq.~\eqref{eq:H_QND}, an ancilla ion $j=0$
is trapped at the edge of the ion chain and is subjected to sideband
resolved laser cooling. The fluorescence emitted by the ancilla ion
is collected by a lens setup and is continuously detected by a homodyne
apparatus. We assume the MS lasers doesn't interact with the ancilla
ion, nor does the cooling laser impact the ions $j\in\{1, \dotsc, N\}$. As such,
the ancilla ion participates in the collective vibrations of the ion
chain and serves as a `transducer' to couple light and phonons, thus
allowing for monitoring the latter.

In the following, we introduce the quantum optical model for our considered
setup in Supplementary Note~\ref{sec:QSSE}, using the language of a quantum stochastic
Schr\"odinger equation (QSSE) (see, e.g., Chap. 9 in Ref.~\cite{gardiner2015quantum}
for an introduction). Based on it, in Supplementary Note~\ref{sec:adiabatic elimination}
we derive a QSSE describing the coupling between the phonons and light
by adiabatically eliminating the internal DOFs of the ancilla ion.
Finally, in Supplementary Note~\ref{sec:SME} we derive the stochastic master equation
for continuous homodyne detection of the spontaneously emitted light
and arrive at Eqs.~\eqref{eq:SME_measureH} and \eqref{eq:Ih_measureH}.

\subsubsection{Quantum stochastic Schr\"odinger equation}

\label{sec:QSSE} To be specific, we consider a standing-wave cooling
configuration, i.e., the ancilla ion locates at the node of the standing
wave~\cite{cirac1992}. In the interaction picture with respect to $\hat{H}_0$ [c.f. Eq.~\eqref{eq: phonon_free_H}], and in the frame rotating with the frequency
of the cooling laser $\omega_{L}$, the internal dynamics of the auxiliary
ion is described by 
\begin{equation}
\hat{H}_{\text{TLS}}=-\Delta_{e}|e\rangle\langle e|+\frac{\Omega_{0}}{2}(|e\rangle\langle g|+{\rm H.c.})\sin[k_{0}\hat{X}_{0}(t)].
\end{equation}
Here, $|g\rangle(|e\rangle)$ is the ground(excited) level of the
cooling transition respectively and $\Delta_{e}=\omega_{L}-\omega_{eg}$
is the frequency detuning between the cooling laser and the $|g\rangle\to|e\rangle$
transition. We assume the cooling laser is along the $x$ axis, with
wavevector $k_{0}$ and Rabi frequency $\Omega_{0}$. The operator
$\hat{X}_{0}(t)$ describes the (small-amplitude) displacement of the
ancilla ion around its equilibrium position, and is related to the
collective phonon modes of the ion chain by $\hat{X}_{0}(t)=\sum_{q}M_{0q}[\hat{a}_{q}(t)+\hat{a}_{q}^\dag(t)]/\sqrt{2m_{0}\omega_{q}}$
with $m_{0}$ the mass of the ancilla ion, and $\hat{a}_{q}(t)=\hat{a}_q{\rm exp}(-i\omega_q t)$.

Besides the internal structure of the ancilla ion, the rest DOFs of
our model includes the internal pseudo-spins of ion $j\in\{1, \dotsc, N\}$ and
the $N+1$ axial phonon modes. In the interaction picture with respect to $\hat{H}_0$, the time evolution of the total system is described by the (It\^o) QSSE~\cite{gardiner2015quantum}
for the ions and the external electromagnetic field (bath DOFs), 
\begin{align}
d|\Psi\rangle= & -i\left(\hat{H}_{\cal{SM}}+\hat{H}_{\text{TLS}}-\frac{i}{2}\Gamma_{e}|e\rangle\langle e|\right)|\Psi\rangle dt\nonumber \\
 & +\int du\sqrt{\Gamma_{e}N(u)}|g\rangle\langle e|e^{-ik_{0}u\hat{X}_{0}(t)}d\hat{B}^{\dagger}(u,t)|\Psi\rangle.\label{eq:QSSE}
\end{align}
In Eq.~\eqref{eq:QSSE}, the first line includes the spin-phonon
Hamiltonian $\hat{H}_{{\rm sys}}$, the internal Hamiltonian of the
ancilla ion $\hat{H}_{{\rm TLS}}$, and the spontaneous decay of the
ancilla ion at a rate $\Gamma_{e}$. The second line describes spontaneous
emission of the ancilla ion into the 3D electromagnetic modes. Here,
the function $N(u)$ reflects the dipole emission pattern of the cooling
transition, which, for the 1D ionic motion considered here, depends
on a single variable $u\equiv\cos\nu\in[-1,1]$ with $\nu$ the angle
between the wavevector of the emitted photon and the $x$ axis. The
spontaneous emission is accompanied by the momentum recoil described
by the operator $e^{-ik_{0}u\hat{X}_{0}}$, with $k_{0}$ the wavevector
of the emitted photon (approximately the same as the wavevector of
the cooling laser). To account for the relevant electromagnetic modes
in the emission direction $u$, quantum optics introduces the corresponding
bosonic noise operators $\hat{b}_{u}(t)$ and $\hat{b}_{u}^{\dag}(t)$,
satisfying the white-noise commutation relations $[\hat{b}_{u}(t),\hat{b}_{u}^{\dag}(t')]=\delta(u-u')\delta({t-t'})$~\cite{gardiner2015quantum}.
In the It\^o QSSE~\eqref{eq:QSSE} these noise operators are transcribed
as Wiener operator noise increments, $\hat{b}_{u}(t)dt\to d\hat{B}(u,t)$.
Assuming the 3D bath is initially in the vacuum state, they obey the
It\^o table~\cite{gardiner2015quantum}, 
\begin{equation}
\begin{split}d\hat{B}(u,t)d\hat{B}^{\dag}(u',t) & =dt\delta(u-u'),\\
d\hat{B}^{\dag}(u,t)d\hat{B}(u',t) & =0,\\
d\hat{B}(u,t)d\hat{B}(u',t) & =d\hat{B}^{\dag}(u,t)d\hat{B}^{\dag}(u',t)=0.
\end{split}
\label{eq:vacuumItoLaw2}
\end{equation}
We note, apart from the explicit ion-bath coupling in the second line
of Eq.~\eqref{eq:QSSE}, the inclusion of the 3D electromagnetic
field bath also introduces a decay term $-i\Gamma_{e}|e\rangle\langle e|/2$
in the first line of Eq.~\eqref{eq:QSSE}. Mathematically, this non-Hermitian
term appears as an ``It\^o correction" when applying the It\^o stochastic
calculus to describe physical systems~\cite{gardiner2015quantum}.

Based on Eq.~\eqref{eq:QSSE}, in the next section we derive a QSSE
describing the coupling between the phonon modes and the electromagnetic
field bath by adiabatically eliminating the internal dynamics of the
ancilla ion.

\subsubsection{Adiabatic elimination of the internal dynamics of the ancilla ion}

\label{sec:adiabatic elimination} We consider the following parameter
regime. (i) The ancilla ion is weakly excited by the cooling laser,
$\eta_{q}^{0}\Omega_{0}\ll\Gamma_{e}$, where $\eta_{q}^{0}\equiv k_{0}/\sqrt{2m_{0}\omega_{q}}$
is the Lamb-Dicke parameter corresponding to the cooling laser. (ii)
The QND interaction is much weaker than the spontaneous emission strength
of the ancilla ion, $|\hat{H}_{{\cal SM}}|\ll\Gamma_{e}$. (iii) The
sideband resolved regime $\omega_{q}\gg\Gamma_{e}$. Condition (i)
and (ii) guarantees that the internal dynamics of the ancilla ion
is much faster than the dynamics of the rest of the system, allowing
us to adiabatically eliminate the internal dynamics of the ancilla
ion. Condition (iii) enables us to selectively enhance the center-of-mass
phonon contribution in the detected photon current (see detailed discussion
in Supplementary Note~\ref{sec:SME}).

To perform the adiabatic elimination, we formally decompose the state
of the total system {[}see Eq.~\eqref{eq:QSSE}{]} into two components,
$|\Psi\rangle=|\psi_{e}\rangle|e\rangle+|\psi_{g}\rangle|g\rangle$,
with $|\psi_{e(g)}\rangle\equiv\langle e(g)|\Psi\rangle$. By the
expansion up to second order in the small Lamb-Dicke parameter $\eta_{q}^{0}$,
Eq.~\eqref{eq:QSSE} becomes two coupled equations for $|\psi_{e(g)}\rangle$,
\begin{align}
d|\psi_{e}\rangle= & -i\left[\hat{H}_{{\cal SM}}-\left(\Delta_{e}+\frac{i}{2}\Gamma_{e}\right)\right]|\psi_{e}\rangle dt\nonumber \\
 & -i\frac{\Omega_{0}}{2}\sum_{q}\eta_{q}^{0}M_{0q}\left[\hat{a}_{q}^{\dagger}(t)+\hat{a}_{q}(t)\right]|\psi_{g}\rangle dt,\label{eq:elimination_psi_e}\\
d|\psi_{g}\rangle= & -i\hat{H}_{\cal SM}|\psi_{g}\rangle dt-i\frac{\Omega_{0}}{2}\sum_{q}\eta_{q}^{0}M_{0q}\left[\hat{a}_{q}^{\dagger}(t)+\hat{a}_{q}(t)\right]|\psi_{e}\rangle dt\nonumber \\
 & +\int du\sqrt{\Gamma_{e}N(u)}\bigg\{1-i\sum_{q}\eta_{q}^{0}M_{0q}\left[\hat{a}_{q}^{\dagger}(t)+\hat{a}_{q}(t)\right]\nonumber \\
 & -\frac{1}{2}\Big[\sum_{q}\eta_{q}^{0}M_{0q}\left[\hat{a}_{q}^{\dagger}(t)+\hat{a}_{q}(t)\right]\Big]^{2}\bigg\}d\hat{B}^{\dagger}(u,t)|\psi_{e}\rangle.\label{eq:elimination_psi_g}
\end{align}
From Eq.~\eqref{eq:elimination_psi_e} it is easy to see $|{\psi}_{e}\rangle\sim O(\eta_{q}^{0})$.
To keep $|\psi_{g}\rangle$ accurate to $O[(\eta_{q}^{0})^{2}]$,
we can neglect the second order Taylor expansion in the last term
of Eq.~\eqref{eq:elimination_psi_g}.

Under conditions (i) and (ii) introduced in the beginning of this
section, Eq.~\eqref{eq:elimination_psi_e} can be solved adiabatically
\begin{align*}
|{\psi}_{e}\rangle=\frac{\Omega_{0}}{2}\sum_{q}\eta_{q}^{0}M_{0q}\bigg[\frac{\hat{a}_{q}^{\dagger}(t)}{\Delta_{e}\!-\!\omega_{q}\!+\!\frac{i}{2}\Gamma_{\!e}}\!+\!\frac{\hat{a}_{q}(t)}{\Delta_{e}\!+\!\omega_{q}\!+\!\frac{i}{2}\Gamma_{\!e}}\bigg]|{\psi}_{g}\rangle.
\end{align*}
Plugging the solution into Eq.~\eqref{eq:elimination_psi_g}, we
arrive at a QSSE which describes the slow dynamics of the system assuming
the ancilla ion staying in its internal stationary (ground) state,
\begin{align}
d|\psi_{g}\rangle= & -i\left(\hat{{H}}_{{\cal SM}}+\sum_{q}\delta{\omega}_{q}\hat{a}_{q}^{\dagger}\hat{a}_{q}\right) |\psi_{g}\rangle dt\nonumber \\
&-\frac{1}{2}\left(A_{q}^{+}\hat{a}_{q}\hat{a}_{q}^{\dagger}+A_{q}^{-}\hat{a}_{q}^{\dagger}\hat{a}_{q}\right)|\psi_{g}\rangle dt\nonumber \\
 & +\int du\sqrt{\Gamma_{e}N(u)}\hat{{\cal J}}d\hat{B}^{\dagger}(u,t)|\psi_{g}\rangle,\label{eq:SSE_phonon}
\end{align}
where $dt\gg1/\Gamma_{e}$ is the coarse-grained time increment and
$d\hat{B}^{\dag}(u,t)$ is the corresponding coarse-grained quantum
noise increment. $\delta\omega_{q}$
is a (tiny) frequency renormalization of the $q$-th phonon mode, 
\begin{align*}
\delta\omega_{q}=(\eta_{q}^{0}M_{0q}\Omega_{0})^{2}\bigg[\frac{\Delta_{e}+\omega_{q}}{4(\Delta_{e}\!+\!\omega_{q})^{2}\!+\!\Gamma_{e}^{2}}+\frac{\Delta_{e}-\omega_{q}}{4(\Delta_{e}\!-\!\omega_{q})^{2}\!+\!\Gamma_{e}^{2}}\bigg].
\end{align*}
In the following we neglect such a tiny frequency shift. The damping rates $A_{q}^{\pm}$
for the $q$-th phonon mode are defined as 
\begin{equation}
A_{q}^{\pm}=\frac{(\eta_{q}^{0}M_{0q}\Omega_{0})^{2}}{4(\Delta_{e}\mp\omega_{q})^{2}+\Gamma_{e}^{2}}\Gamma_{e}.
\end{equation}
The operator $\hat{{\cal J}}$ is a collective quantum jump operator
including all phonon modes, 
\begin{align}
\hat{{\cal J}}=\frac{\Omega_{0}}{2}\sum_{q}\eta_{q}^{0}M_{0q}\bigg(\frac{\hat{a}_{q}^{\dagger}(t)}{\Delta_{e}-\omega_{q}+\frac{i}{2}\Gamma_{e}}+\frac{\hat{a}_{q}(t)}{\Delta_{e}+\omega_{q}+\frac{i}{2}\Gamma_{e}}\bigg).\label{eq:collapse_OP_full}
\end{align}

The QSSE~\eqref{eq:SSE_phonon} describes the coupling between the
phonon DOFs and the external electromagnetic field bath. This allows
us to read out the COM quadrature $\hat{X}$ via homodyne detection
of the external bath, as detailed in the next section. 

\subsubsection{Homodyne detection of the fluorescence}

\label{sec:SME} We consider continuous homodyne detection of the
laser cooling fluorescence, as shown schematically in Fig.~1 of the
main text. In such a measurement, the fluorescence photons are collected
by linear optical elements, e.g., by a lens setup, and are then mixed
with a reference laser at a beam splitter. Photon counting of the
mixed beam then allows for the measurement of the phase information
of the fluorescence photons.

We assume the lens system covers a solid angle $\Omega$, and define
\begin{equation}
\epsilon=\int_{\Omega}duN(u)
\end{equation}
as the fraction of photons collected by the lens setup. The corresponding
quantum noise increment is 
\begin{equation}
d\hat{B}(t)=\frac{1}{\sqrt{\epsilon}}\int_{\Omega}du\sqrt{N(u)}d\hat{B}(u,t).
\end{equation}

The homodyne measurement corresponds to making a measurement of the
following quadrature operator~\cite{wiseman2009quantum,gardiner2015quantum}
\begin{equation}
d\hat{Q}(t)=d\hat{B}(t)e^{-{i\phi}}+d\hat{B}^{\dagger}(t)e^{{i\phi}},
\end{equation}
with $\phi=(\omega_{\rm LO}-\omega_L)t+\phi_0$ and $\omega_{\rm LO}$ and $\phi_0$ being the frequency and phase of the local oscillator. The measurement projects the
state of the bath onto an eigenstate of $d\hat{Q}(t)$ corresponding to the
eigenvalue $dq(t)$, which defines the homodyne current via $dq(t)\equiv
I(t)dt$.
It can be shown~\cite{wiseman2009quantum,gardiner2015quantum} that the
measurement outcome $dq(t)$ obeys a normal distribution centered at the mean
value of the quantum jump operator $\hat{{\cal J}}$, i.e.,
\begin{equation}
dq(t)\equiv I(t)dt=\sqrt{\epsilon\Gamma_{e}}\langle\hat{{\cal J}}e^{-i\phi}+\hat{{\cal J}}^{\dag}e^{i\phi}\rangle_{c}+dW(t),\label{eq:Ih_full}
\end{equation}
where $dW(t)$ is a random Wiener increment, which is related to the
shot noise by $dW(t)=\xi(t)dt$. The expectation value $\langle\dots\rangle_{c}={\rm Tr}(\dots\mu_{c})$
is taken with a conditional density matrix $\mu_{c}$ of the spin-phonon
system. The evolution of $\mu_{c}$ is given by a SME derived from
Eq.~\eqref{eq:SSE_phonon} by projecting out the bath DOFs following
the standard procedure~\cite{wiseman2009quantum,gardiner2015quantum},
\begin{align}
d\mu_{c}= & -i[\hat{{H}}_{{\cal SM}},\mu_{c}]dt\nonumber +\sum_{q}\left(A_{q}^{+}\mathcal{D}[\hat{a}_{q}^{\dagger}]+A_{q}^{-}\mathcal{D}[\hat{a}_{q}]\right)\mu_{c}dt\nonumber \\
 & +\sqrt{\epsilon\Gamma_{e}}\mathcal{H}[\hat{{\cal J}}e^{-i\phi}]\mu_{c}dW(t),\label{eq:SME_app}
\end{align}
with $\mathcal{D}[\hat{O}]\rho\equiv\hat{O}\rho\hat{O}^{\dagger}-\frac{1}{2}\hat{O}^{\dagger}\hat{O}\rho-\frac{1}{2}\rho\hat{O}^{\dagger}\hat{O}$
being the Lindblad superoperator, and ${\cal H}[\hat{O}]\rho\equiv\hat{O}\rho-{\rm Tr}(\hat{O}\rho)\rho+\text{H.c.}$
a superoperator corresponding to homodyne measurement. The first two
lines of Eq.~\eqref{eq:SME_app} is akin to the laser cooling master
equation of trapped particles~\cite{cirac1992,gardiner2015quantum},
while the third line describes the measurement backaction of a continuous
homodyne detection.

Under the condition of resolved sideband $\omega_{q}\gg\Gamma_{e}$,
we can enhance the component corresponding to the COM phonon mode
in the homodyne signal Eq.~\eqref{eq:Ih_full}, by tuning the cooling
laser in resonance with the red sideband of the COM mode, $\Delta_{e}=-\omega_{0}$.
Under this condition, we have $\hat{{\cal J}}\simeq-i\Omega_{0}\eta_{0}^{0}M_{00}\hat{a}_{0}{\rm exp}(-i\omega_0 t)/\Gamma_{e}$
{[}see Eq.~\eqref{eq:collapse_OP_full}{]}, and $A_{0}^{+}\simeq A_{q}^{\pm}\simeq0$
for $q\neq0$. Defining $\hat{\rho}_{c}^{{\cal SM}}={\rm Tr}_{{\rm ph},q\neq0}({\mu}_{c})$
by trancing out the phonon modes except for the COM mode, we have
\begin{align}
I(t)= & \sqrt{2\epsilon\gamma_{s}}\langle\hat{X}\rangle_{c}+\xi(t),\nonumber \\
d\rho_{c}^{{\cal SM}}= & -i[\hat{{H}}_{{\cal SM}},\rho_{c}^{{\cal SM}}]dt+\gamma_{s}\mathcal{D}\left[\hat{a}_{0}\right]\rho_{c}^{{\cal SM}}dt\nonumber \\
 & +\sqrt{\epsilon\gamma_{s}}\mathcal{H}\left[\hat{a}_{0}\right]\rho_{c}^{{\cal SM}}dW(t).\label{eq:SME_eff}
\end{align}
where $\hat{X}=(\hat{a}_{0}+\hat{a}_{0}^{\dag})/\sqrt{2}$ is the
$x$-quadrature of the COM phonon mode, $\gamma_{s}=(\Omega_{0}\eta_{0}^{0}M_{00})^{2}/\Gamma_{e}$
is an effective measurement rate, with $M_{00}\simeq1/\sqrt{N}$,
and we choose $\omega_{\rm LO}-\omega_L=\omega_0$ and $\phi_0=-\pi/2$ for the local oscillator to maximize the homodyne current.

Equation \eqref{eq:SME_eff} already describes continuous QND readout
of the transverse field Ising Hamiltonian. To simplify the analysis,
we can further adiabatically eliminating the COM phonon mode in Eq.~\eqref{eq:SME_eff}
under the condition $\gamma_{s}\gg\vartheta J$, and arrive at Eqs.~\eqref{eq:SME_measureH}
and ~\eqref{eq:Ih_measureH} with the identification $\gamma\equiv2J^{2}\vartheta^{2}/\gamma_{s}=2\Gamma_{e}(\vartheta J/\Omega_{0}\eta_{0}^{0}M_{00})^{2}$.

\subsubsection{Filtering of the homodyne current}

The homodyne current Eq.~\eqref{eq:Ih_measureH} is noisy, as it
contains the (white) shot noise $\xi(t)$ inherited from the vacuum
fluctuation of the electromagnetic field environment. To suppress
the noise, we filter the homodyne current with a suitable linear lowpass
filter 
\begin{equation}
\mathcal{I}_{\tau}(t)=\int dt'h_{\tau}(t-t')I(t'),\label{eq:filteredIh}
\end{equation}
where $h_{\tau}(t)$ is the filter function with a frequency bandwidth
$\sim1/\tau$, and $\mathcal{I}(t)$ is the \emph{filtered homodyne
current}. The filter attenuates the component of the shot noise with
frequency higher than $1/\tau$ thus allowing us to extract out the
signal we are interested in.

We adopt two filters in the main text. The first one is a simple \emph{cumulative
time-average}, $\overline{\mathcal{I}}(\tau)=(2N\sqrt{\gamma\epsilon\tau})^{-1}\int_{0}^{\tau}dtI(t)$.
This allows us to attenuate the shot noise as much as possible, and
is especially suitable for QND measurement (cf. Fig. 1e of the main
text). In contrast, for imperfect QND measurement we are interested
in resolving the quantum jumps between different energy eigenstates
as a competition between coherent evolution and measurement backaction.
To achieve this, we filter the homodyne current via $\mathcal{I}_{\tau}(t)=(2N\sqrt{\gamma\epsilon}\tau)^{-1}\int_{0}^{\infty}dt'e^{-t'/\tau}I(t-t')$
and call $\mathcal{I}_{\tau}(t)$ the \emph{window-filtered homodyne
current}. The time window $\tau$ is chosen to ensure $1/\gamma\ll\tau\ll T_{{\rm dwell}}$
with $\gamma$ the measurement rate and $T_{{\rm dwell}}$ the typical
time that the system dwells in particular eigenstates. This allows
us to attenuate the shot noise as much as possible while still being
able to resolve the quantum jumps.

\subsection{Experimental feasibility}
\label{sec:experiment}

Having discussed our QND measurement scheme for the transverse-field Ising
Hamiltonian in trapped-ion setups, in this section, we show that
state-of-the-art trapped-ion experiments provide all ingredients for the
implementation of the QND scheme. First, in Supplementary Note~\ref{sec:exp_general}, we
summarize the experimental requirements of our scheme and discuss experimental
imperfections including multiple sources of decoherence. We then discuss some
practical points. These include the implementation of our scheme with axial and
transverse phonon modes, analyzed in Supplementary Note~\ref{sec:axial_phonon} and
Supplementary Note~\ref{sec:transverse_phonon} respectively, as well as the implementation
with different ion species, discussed in Supplementary Note~\ref{sec:Ion_species}. Finally, in
Supplementary Note~\ref{sec:numbers}, we present experimental parameters for
proof-of-principle realizations of our scheme.

\subsubsection{Experimental requirements and practical imperfections}
\label{sec:exp_general}

The performance of our QND measurement scheme depends on the collection
efficiency $\epsilon$ of the photons scattered by the ancilla ion. A collection
efficiency of $15 \, \%$ is experimentally feasible for a single trapped
ion~\cite{PhysRevLett.96.043003}, and we expect that a similar collection
efficiency can be reached in our proposed setup. Even larger photon collection
rates can be achieved by coupling the ancilla ion to optical
cavities~\cite{Stute2012}, or by simultaneous detection of the fluorescence of
several ancilla ions.

In the implementation of homodyne detection of the spin system, we assume that
the MS lasers do not interact with the ancilla ion, and that the cooling laser
does not impact the ions $j\in\{1, \dotsc, N\}$. These requirements can be met by
individual addressing of each ion in realizing the MS
configuration. Alternatively, this can be achieved by using global MS lasers and
by choosing the ancilla ion from a different ion
species~\cite{Tan:2015aa,Negnevitsky:2018aa}, so that the ancilla is decoupled
from the MS lasers due to its different internal electronic structure. We note,
however, that an ancilla ion with a different mass changes the structure of the
COM mode. This has to be rectified in order to perform our QND measurement
scheme, e.g., via local adjustments of the trapping potential near the ancilla
ion using optical potentials~\cite{Schneider:2010aa}.

Realistic trapped-ion systems have multiple sources of decoherence. The
coherence time of current trapped-ion quantum simulators is limited by dephasing
of the internal spins due to fluctuations of the global magnetic field which
defines the quantization axis. Encoding the spins in ionic internal states which
are first-order insensitive to magnetic field fluctuations greatly suppresses
dephasing and extends the single-spin coherence time. This has been implemented,
e.g., for $^9{\rm Be}^+$ ions (with a {single-spin coherence time $t_{\rm coh}$}
$\sim 1.5 \, {\rm s}$~\cite{PhysRevLett.95.060502}) and for $^{171}{\rm Yb}^+$
ions (with {$t_{\rm coh}\sim 2.5 \, {\rm s}$}~\cite{PhysRevA.76.052314}).
Without this type of encoding, the coherence time is typically one order of
magnitude shorter. For example, {for $^{40}{\rm Ca}^+$ ions $t_{\rm coh}\sim 95 \, \mathrm{ms}$}~\cite{PhysRevLett.106.130506}.

Another important source of decoherence is phonon heating due to electromagnetic
field noise. In standard linear Paul traps, the phonon heating rate is typically
below $1/$s and is thus negligible~\cite{PhysRevLett.83.4713}. Nevertheless, in
surface ion traps, phonon heating is much more significant due to the short
distance between the ions and the trap electrodes. Operating at cryogenic
temperature can reduce phonon heating significantly. For example, the phonon
heating rate of axial phonons is reduced to values as low as $70/$s for ion
spacings of $d\sim30 \, \mu \mathrm{m}$ in the cryogenic surface traps which are
used by the NIST group~\cite{Brown2011}. Even lower phonon heating rates are
being actively pursued~\cite{McConnell2015}.

In Supplementary Notes~\ref{sec:Ion_species} and~\ref{sec:numbers} further below, we show that
the proposed QND measurement requires a time much shorter than the coherence
time of trapped ions which is limited by the factors outlined above. Thus,
current trapped-ion technology allows for robust implementation of our QND
measurement scheme.

\subsubsection{Implementation with axial phonon modes}
\label{sec:axial_phonon}

In this section, we present some considerations on implementation of our QND
measurement scheme with axial phonon modes.

The spectrum of axial phonon modes of an
ion string in a linear Paul trap is extensive, i.e., it broadens with
increasing number of ions $N$. To implement the long range Ising
model with dipolar interaction $J_{ij}\propto 1/|i-j|^3$, the detunings $\Delta(\Delta')$ of the double MS configuration should
also increase with the number of ions. Thus, to keep the spin-spin
coupling $J\propto(\Omega/\Delta)^{2}(k^{2}/2m)$ {[}see Eq.~\eqref{eq:J_mn}{]}
finite, the power of the MS laser beams also goes up with increasing
$N$. The achievable laser power in the laboratory thus puts a practical
limitation on the scalability of the implementation wit axial phonon
modes. On the other hand, the implementation with axial phonon modes benefits
a relative large system-meter coupling $\hat{H}_{{\cal SM}}$, thanks to the large Lamb-Dicke parameter $\eta$ associated with axial
phonon modes (we note that in $\hat{H}_{{\cal SM}}$, $\vartheta\propto\eta$).
In view of these, the implementation with axial phonon modes best serves
as a small-scale proof-of-principle experiment, which demonstrates our proposed
QND measurement and its applications. We provide the typical experimental parameters
for such an implementation in Supplementary Note~\ref{sec:numbers}.

Moreover, we comment that the extensive feature of the axial phonon spectrum allows for engineering exotic spin coupling that goes beyond the power-law coupling, e.g., frustrated spin models, by carefully adjusting the laser detunings $\Delta^{(\prime)}$ with respect to the phonon spectrum~\cite{Porras2004}. The associated QND measurement could possibly enable rich opportunities for the study of these models and the preparation of their eigenstates.

\subsubsection{Implementation with transverse phonon modes}

\label{sec:transverse_phonon} 
Here we present some considerations concerning the implementation of our QND
measurement scheme with transverse phonon modes.

In contrast to axial phonon modes, the transverse phonon modes in a linear Paul
trap have a dense spectrum of width $\propto\omega_{z}^{2}/\omega_{x}$, which is
almost independent of the number of ions $N$; Here, $\omega_{z(x)}$ is the
trapping frequency along the axial (transverse) direction, respectively. As a
result, the long range-Ising model and the associated QND measurement can be
implemented by a double MS configuration for which the detunings
$\Delta^{(\prime)}$ and the Rabi frequency $\Omega$ can be held fixed upon
increasing the number of ions. This leads to better scalability regarding the
laser power as compared to the implementation with axial phonon modes.

The scalability of an implementation with transverse phonon modes is limited by
the condition Eq.~\eqref{eq:validity}, since longer ion chain leads to denser
phonon spectrum which eventually violates Eq.~\eqref{eq:validity}. To estimate an
upper limit of the ion number $N$, we note that the LHS of
Eq.~\eqref{eq:validity} is much smaller than $\sum_q\eta_q^2\Omega^2\eta/|\Delta^{(\prime)}-\omega_0|\ll \sum_q\eta_q^2\Omega$, the latter ``$\ll$" coming from our off-resonance condition $|\Delta^{(\prime)}-\omega_0|\gg\eta\Omega$. Thus, Eq.~\eqref{eq:validity} is well satisfied as long as $\sum_q\eta_q^2\Omega\leq |\omega_1-\omega_0|$. We can estimate $\sum_q\eta_q^2\Omega$ as $N\eta^2\Omega$ and  $|\omega_1-\omega_0|$ as $\omega_z^2/\omega_x$, thus the above condition becomes $N\eta^2\Omega\leq\omega_z^2/\omega_x$.
Moreover, to prevent zig-zag transition of a linear ion chain we
require $\omega_{x}/\omega_{z}\geq0.73N^{0.86}$~\cite{Wineland1998}.  Combining
these two conditions we find $N\leq[\omega_{z}/(\eta^{2}\Omega)]^{0.54}$.  The
latter quantity is typically around $100$ in experiments. Thus, according to
these estimates, the implementation with transverse phonon modes allows for
scaling up to hundreds of ions.

\subsubsection{Implementation with different ion species}
\label{sec:Ion_species}
\begin{figure}
  \includegraphics[width=1\columnwidth]{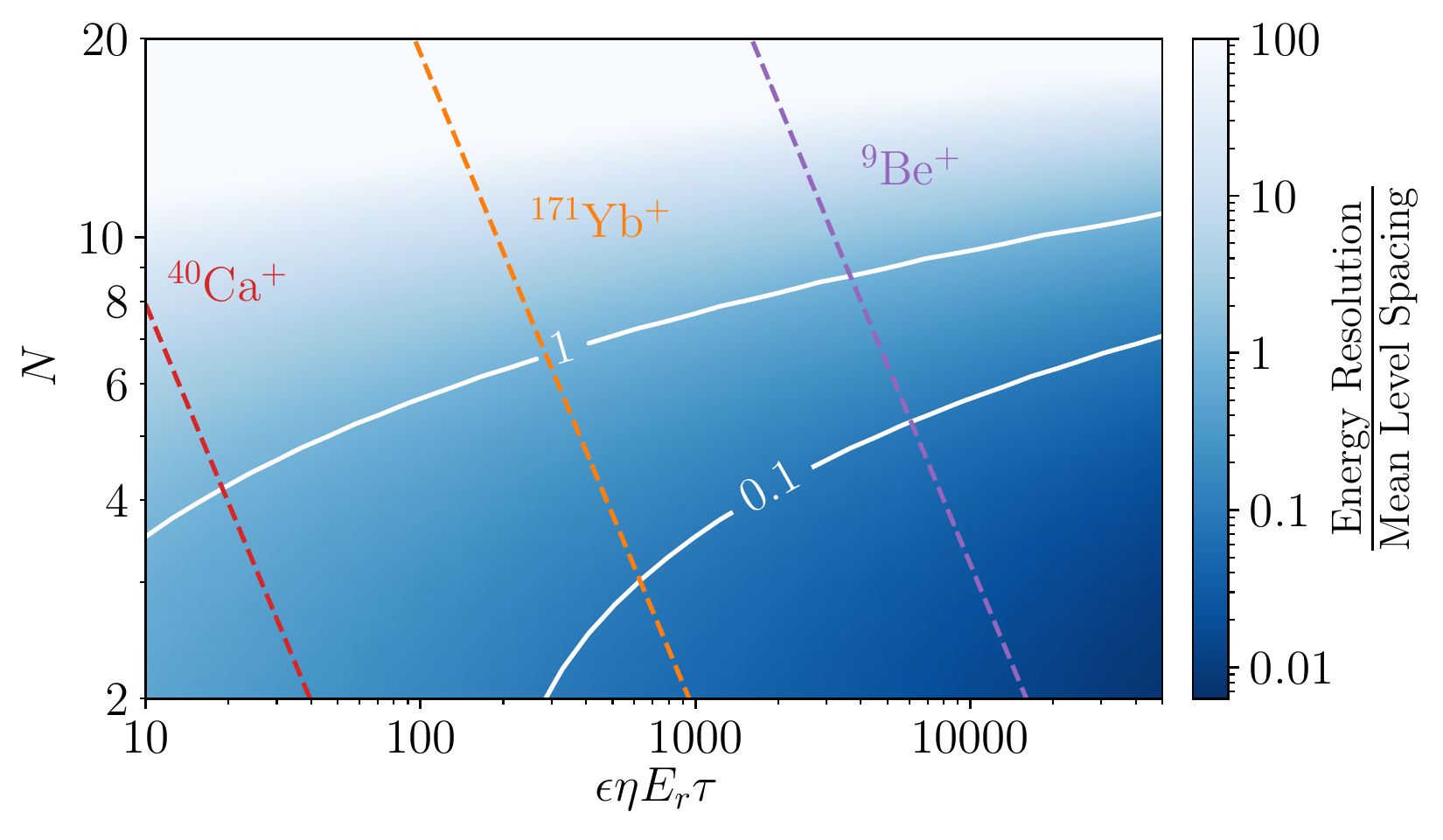}\caption{Energy
    resolution of a single run of the QND measurement for different ions
    species. The (dimensionless) energy resolution $\Delta E/J$ over the mean
    level spacing $N/2^N$ is plotted against the ion number $N$ and the
    dimensionless parameter $\epsilon \eta E_r\tau$, where $\epsilon$ is the
    detection efficency, $\eta$ the Lamb-Dicke paramter $E_r$ the recoil energy
    and $\tau$ the measurement time. {The three dashed lines corresponds to
    estimations for three different ion species, for which we fix
    $\epsilon=0.15, \eta=0.1$ and choose $\tau=t_{\rm coh}/N$ with $t_{\rm coh}$ the single-spin coherence time. }}
\label{Fig:IonSpecies}
\end{figure}
We now discuss and compare the implementation with different ion species. As
already mentioned in Supplementary Note~\ref{sec:exp_general}, different ion species have
different coupling strengths to the laser fields, and thus different energy
scales of the targeted spin models and the associated QND measurements. Further,
different ion species possess different coherence time. These parameters impact
the performance of an implementation of our QND measurement scheme. As examples,
we consider three ion species that are commonly used in current trapped-ion
experiments: $^{171}{\rm Yb}^+$, $^{40}{\rm Ca}^+$ and $^{9}{\rm Be}^+$.

A key parameter to quantify the performance of the proposed QND measurement is
the signal-to-noise ratio (SNR) of a single measurement run, see the discussion
in Methods of the main text. This can be equivalently expressed in terms of the
(dimensionless) energy resolution $\Delta E/J\sim 1/\sqrt{2\gamma\epsilon\tau}$,
for given measurement strength $\gamma$, photon collection efficiency $\epsilon$
and filtering time $\tau$ (which is the same as the measurement time). The
energy resolution represents our ability to distinguish two adjacent energy
levels from the documented data of a single measurement run, and should be
compared to the (dimensionless) mean many-body level spacing, which can be
estimated as $N/2^N$.

{The achievable measurement time $\tau$ can be estimated using the typical many-body dephasing time as $\tau= t_{\rm coh}/N$, with $t_{\rm coh}$ the single-spin coherence time and $N$ the number of ions.} The measurement strength $\gamma$
of our QND scheme is controlled by the parameter
$\vartheta J\sim (\eta/\sqrt{N})E_r(\Omega/\Delta)^2$, see the discussion in
Supplementary Note~\ref{subsec:Double-Molmer-Sorensen-interaction}. Here, $E_r$ is the recoil
energy of the qubit transition, whereas $\Omega/\Delta\ll 1$ is a small quantity
independent of the ion species. As a result, we have the energy resolution
{$\Delta E/J\sim (\Delta/\Omega)N^{3/4}/\sqrt{2\eta \epsilon E_r t_{\rm coh}}$}.
Taking realistic experimental parameters, we find that the quantity
$\eta \epsilon E_r t_{\rm coh}$ differs for different ion species and spans a
range from $10^2$ for $^{40}{\rm Ca}^{+}$ to $10^4$ for $^9{\rm Be}^{+}$, see
the vertical lines in Supplementary Fig.~\ref{Fig:IonSpecies}. Among the three ion species,
$^9{\rm Be}^{+}$ holds the promise of achieving the best energy resolution due
to its light mass and long coherence time.

In Supplementary Fig.~\ref{Fig:IonSpecies}, we further plot the ratio between the energy
resolution $\Delta E/J$ and the mean level spacing $N/2^N$, for increasing ion
number $N$. Under current experimental conditions, the maximum system size $N$
for which a single run of the QND measurement is able to resolve the
eigenenergies can be estimated as {$N\simeq 4$ for $^{40}{\rm Ca}^{+}$,
$N\simeq 6$ for $^{171}{\rm Yb}^{+}$ and $N\simeq 8$ for $^{9}{\rm Be}^{+}$}.
As a result, all three ion species are good candidates for building an
intermediate-size interacting spin system for testing quantum fluctuation
relations. The favorable scalability of $^{9}{\rm Be}^{+}$ facilitates testing
the eigenstate thermalization hypothesis. Finally, we note that these
estimations concern a single measurement run and better energy resolution could
be achieved by repeated measurements.

\subsubsection{Parameters for proof-of-principle experiments}
\label{sec:numbers} 
We proceed to present experimental parameters for a proof-of-principle
implementation of our QND measurement scheme.

First, let us consider an implementation with $^{9}$Be$^{+}$ ions and with axial
phonon modes. The experimental system we have in mind is similar to the one
reported in Ref.~\cite{PhysRevLett.117.060505}. To be concrete, we consider
$N=5$ $^{9}$Be$^{+}$ ions in a linear Paul trap. The internal spin of a
$^{9}$Be$^{+}$ ion consists of two hyperfine states driven by a Raman transition
involving two $313 \, \mathrm{nm}$ single-photon transitions with recoil energy
$E_{r}=2\pi\times226.5 \, \mathrm{kHZ}$. We choose the axial trapping frequency
$\omega_{z}=2\pi\times3 \, \mathrm{MHz}$, leading to a moderate Lamb-Dicke
parameter of $\eta\simeq\sqrt{E_{r}/\omega_{z}}\simeq0.27$. We further choose
$\Delta=3\omega_{q=4}$ and $\Omega=0.15\Delta$ to stay in the off-resonant
regime. The resulting spin-spin coupling strength is
$J\simeq2\pi\times5 \, \mathrm{kHz}$, and the system-meter coupling is
$\vartheta\simeq-0.17$. We choose the laser cooling rate of the ancilla ion
$\gamma_{s}=2\pi\times5 \, \mathrm{kHz}$. Consequently, the effective
measurement rate is $\gamma\simeq2\pi\times290 \, \mathrm{Hz}$. Assuming a
photon collection efficiency $\epsilon=0.15$ as discussed above, our QND
measurement has a resulting characteristic time scale
$1/\epsilon\gamma\simeq3.7 \, \mathrm{ms}$, which is much shorter than the
typical single qubit dephasing time
$\sim1 \, \mathrm{s}$~\cite{PhysRevLett.117.060505}. Specifically, an averaging
time $\tau=10/\epsilon\gamma$ leads to an energy resolution (see Methods)
$\Delta E/J\sim0.22$, smaller than the minimal energy gap in this five-spin
Ising model. This enables the preparation of single energy eigenstates via QND
measurement, which suffices, e.g., for testing quantum fluctuation relations. This also allows for
the observation of quantum jumps between different eigenstates in the imperfect
QND regime as discussed in the main text.

Next, we provide experimental parameters for a transverse-phonon implementation realizing the power-law decaying spin-spin interactions, and discuss the associated energy resolution. These are relevant to the discussion in Supplementary Note~\ref{sec:eigenst-therm-hypoth} below on testing the eigenstate thermalization hypothesis. We
consider $N=6$ $^{9}$Be$^{+}$ ions in a linear Paul trap with axial trapping
frequency $\omega_{z}=2\pi\times 2 \, \mathrm{MHz}$ and transverse trapping
frequency $\omega_{x}=2\pi\times 8 \, \mathrm{MHz}$. We choose
$\Omega=2\pi\times 1.76 \, \mathrm{MHz}$,
$\Delta=2\pi\times 8.8 \, \mathrm{MHz}$, $\Delta^\prime=\Delta+\omega_x$, and
the Lamb-Dicke parameter along the transverse direction
$\eta=0.09<\sqrt{E_r/\omega_x}$, which can be realized by properly choosing the
direction of the double MS beams with respect to the ion string. The resulting
spin-spin coupling strength obeys an approximate power law decay
$J_{ij}\sim J/|i-j|^\alpha$ with $\alpha=1.5$ and
$J\simeq2\pi\times 2.6 \, \mathrm{kHz}$.  Further, this generate a QND coupling
with strength $|\vartheta J|\simeq 2\pi\times 0.1 \, \mathrm{kHz}$. We choose
the laser cooling rate of the ancilla ion
$\gamma_{s}=2\pi\times 0.5 \, \mathrm{kHz}$. Consequently, the effective
measurement rate is $\gamma\simeq2\pi\times 40 \, \mathrm{Hz}$.  Assuming a
photon collection efficiency $\epsilon=0.15$, and a measurement time
$\tau=50 \, \mathrm{ms}$, the achieved energy resolution is
$\Delta E/J\simeq 0.35$. This corresponds to a resolution of the energy density
$\Delta\varepsilon=\Delta E/(JN)=0.06$, which is indicated as horizontal error
bars in Supplementary Fig.~\ref{FIG_eth}.

\section{Numerical study of thermal properties of energy eigenstates}
\label{sec:numer-study-therm}

In this section we provide additional details on the numerical simulations used in our study of thermal properties of the energy eigenstates in the main text. 

In  Supplementary Note~\ref{sec: MC} we  describe the canonical-ensemble quantum Monte-Carlo simulations of the transverse-field Ising model used in Fig. 3 of the main text. We discuss the phase
transition in the case of of long- and short-range interactions.  

In Supplementary Note \ref{sec: Realistic} we discuss the phase diagram of the transverse field Ising model for realistic spin-spin interaction $J_{ij}$ and show that it qualitatively agree with the one obtained using an approximate power-law.

\begin{figure}[t!]
\centering{}\includegraphics[width=0.49\columnwidth]{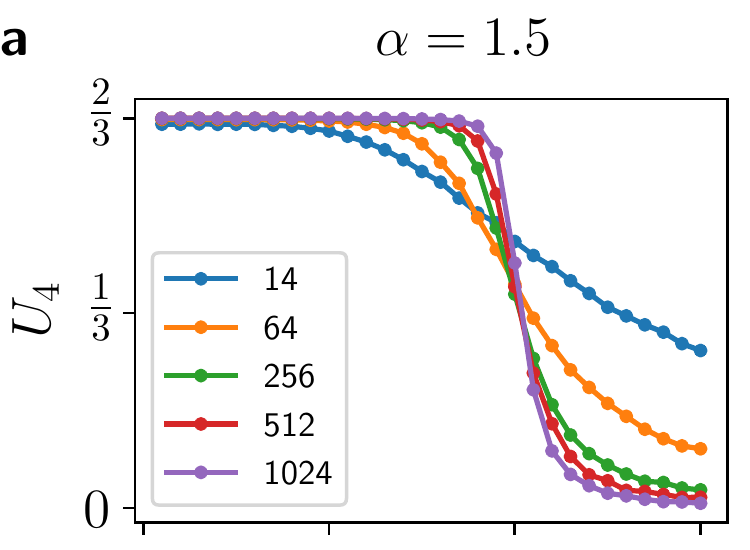}\includegraphics[width=0.49\columnwidth]{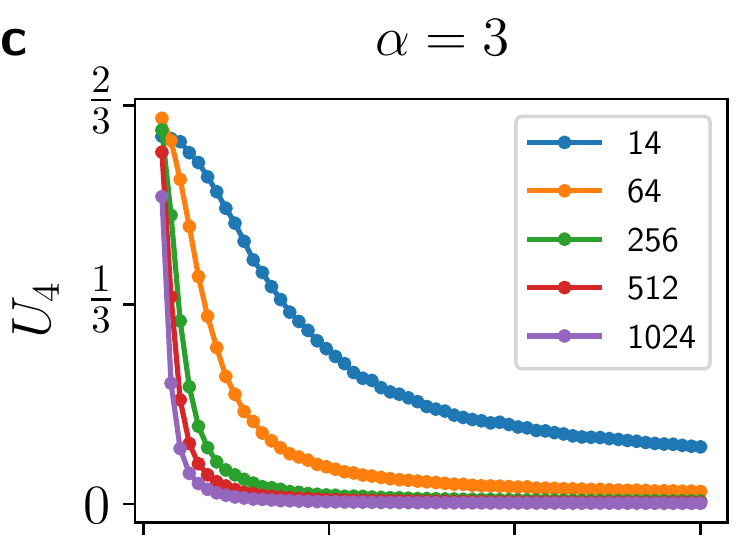}\\
 \includegraphics[width=0.49\columnwidth]{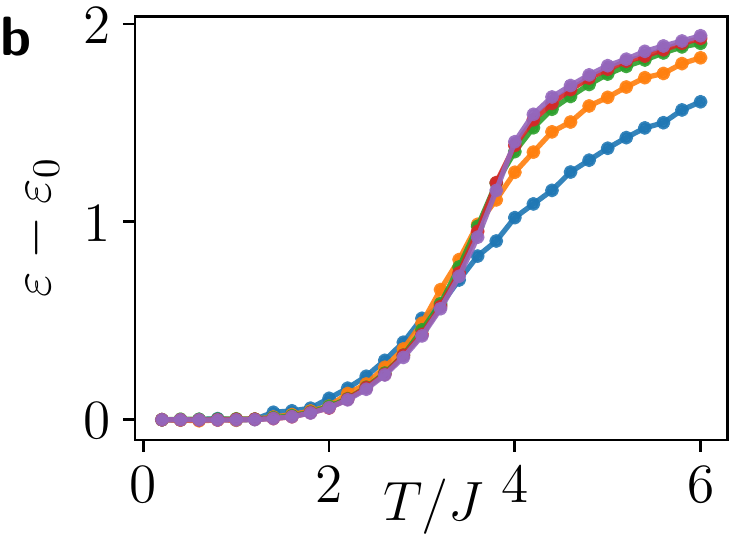}
 \includegraphics[width=0.49\columnwidth]{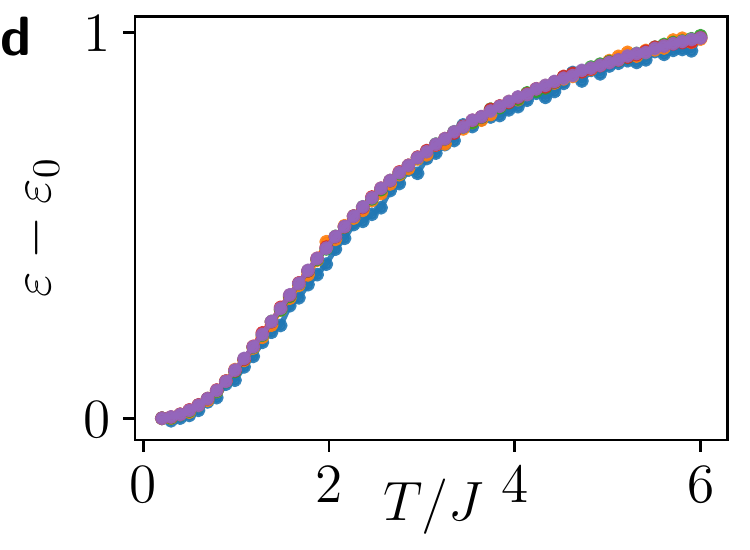}
\caption{Phase transition of the Ising model in the canonical ensemble. (a),
(c) Binder cumulant as a function of temperature for different system
sizes. (b), (d) the corresponding energy of the system.}
\label{Fig1} 
\end{figure}

\subsection{Monte-Carlo simulations}\label{sec: MC}

Here we provide details on the numerical simulations of the phase transition of
the Ising model in canonical ensemble
$\hat{\rho}_{\text{th}}\left(T\right)\equiv
e^{-\hat{H}/T}/\text{Tr}\left[e^{-\hat{H}/T}\right]$
using the quantum Monte-Carlo technique. It allows us to calculate the critical
energy $\varepsilon$ using the finite-size scaling analysis of the Binder
cumulant, defined as
\[
U_{4}\equiv1-\frac{\left\langle \hat{m}_{x}^{4}\right\rangle }{3\left\langle \hat{m}_{x}^{2}\right\rangle ^{2}}
\]
By its construction~\cite{Binder81}, this cumulant distinguishes the ordered
phase with $U_{4}\approx2/3$, from the disordered phase with $U_{4}\approx0$. As
a result, when crossing the phase transition, the Binder cumulant has a sharp
jump between these two values at the critical temperature $T_{c}$. This allows
us to determine $T_{c}$ for the Ising model. The results of the Monte-Carlo
simulation for $\alpha=1.5,$ $h/J=1$ and different system sizes $N$ are shown in
Supplementary Fig.~\ref{Fig1}(a). For a sufficiently large number of spins the curves of
$U_{4}$ cross approximately at the same temperature, which provides a good
estimate of $T_{c}$. The corresponding critical energy density
$\varepsilon\equiv\text{Tr}\left[\hat{H}\hat{\rho}_{\text{th}}\left(T_{c}\right)\right]/NJ$
can be easily determined from the energy-temperature conversion curve shown in
Supplementary Fig.~\ref{Fig1}(b).

\begin{figure}[t!]
\centering{}\includegraphics[width=0.8\columnwidth]{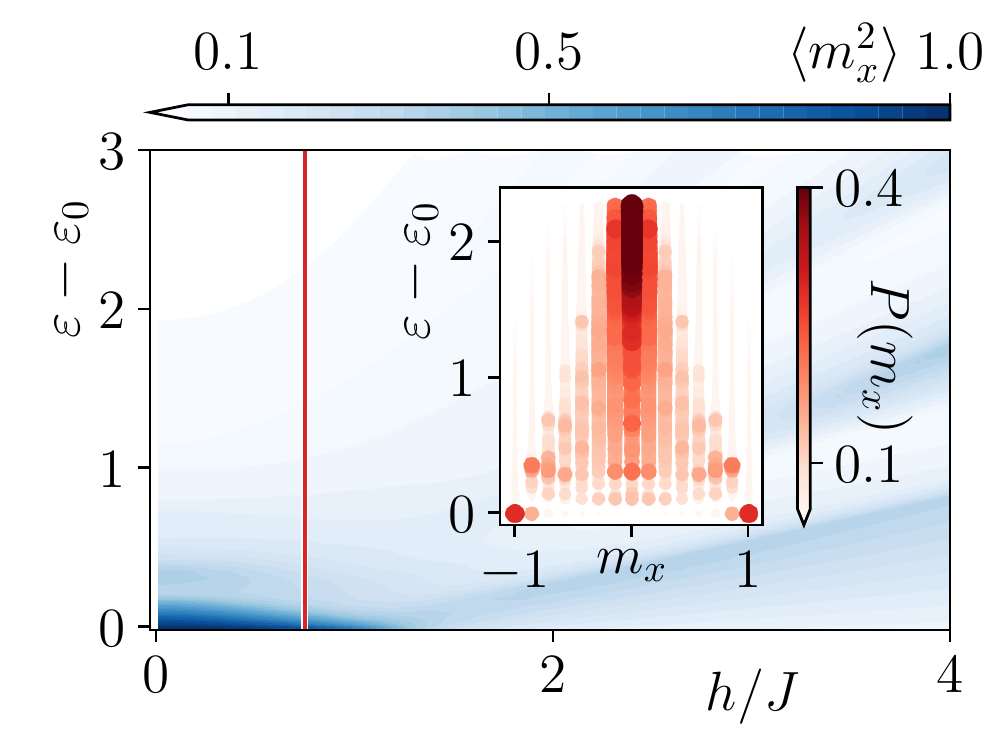}
\caption{Excited-state phase transition in the Ising model with $\alpha=3$.
(a) Ferro-paramagnet crossover in the Ising model of $N=14$ spins
prepared by the energy measurements in microcanonical ensembles of
width $\Delta E/(JN)=0.1$. The transition between magnetically ordered
phase $\braket{\hat{m}_{x}^{2}}_{\text{mc}}\approx1$ (dark blue)
to disordered phase $\braket{\hat{m}_{x}^{2}}_{\text{mc}}\approx0$
(light blue) is shown as function of the mean energy density $\varepsilon=\braket{\hat{H}}_{{\rm mc}}/(JN)$
and the transverse field $h$. Test of ETH for the symmetry sector
$\{+1,+1\}$ is shown in the inset: only the ground state has a bimodal
distribution $P\left(m_{x}\right)$.}
\label{Fig2} 
\end{figure}

We also study the Ising model with $\alpha=3$ shown in Supplementary Fig.~\ref{Fig1}(c) and
(d). The Binder cumulant curves show no crossing at finite temperature, which
indicates the absence of thermal phase transitions as it should be in case of
short-range interactions $\alpha>2$~\cite{Dutta_01}. Below, we study if the same
thermodynamic properties are exhibited by the individual eigenstates as can be
expected if the ETH holds in this regime.

\subsection{Realistic spin-spin interaction}\label{sec: Realistic}
For a finite ion chain, the spin-spin coupling $J_{ij}$ Eq.~\eqref{eq:J_mn} only approximately satisfy the power law. 
Here we study the phase diagram of the Ising model using these realistic spin-spin interaction coefficients $J_{ij}$, and show that it agrees well with the phase diagram of the power-law interaction model studied in the main text. We consider $14$ ions in a linear Paul trap and use the transverse phonons to implement our QND measurement. We choose $\omega_x/\omega_z=10$, $\Delta/\omega_z=10.22$ and $\Delta^\prime/\omega_z=20.22$. By numerically calculate the phonon modes, we find that the spin-spin couplings in Eq. (8) of the main text satisfy approximately power-law decay, with the exponent $\alpha\simeq 1.5$ from a least-square fit. Based on these realistic spin-spin couplings, we calculate the phase diagram of the system as is shown in Supplementary Fig. \ref{Fig_realistic}. As can be seen all the features of the phase diagram are in good agreement with those shown in the main text, where we approximated $J_{ij}\propto1/|i-j|^{1.5}$.

\subsection{Case of short-range interactions}

We now study the phase transition for the case $\alpha=3$ in the
microcanonical ensemble and on the level of individual eigenstates.
The phase diagram in the microcanonical ensemble is shown in Supplementary Fig.~\ref{Fig2}.
It is clearly visible that contrary to the long-range Ising model,
the ordering remains only in the vicinity of $\epsilon\approx\epsilon_{0}$.
This is also reflected by the order parameter probability distribution
$P\left(m_{x}\right)$ for the individual eigenstates shown in inset
of Supplementary Fig.~\ref{Fig2}, which shows bimodal behavior only for the ground
state. We note that this observation is compatible with the eigenstate
thermalization hypothesis.

\begin{figure}[t!]
\centering{}\includegraphics[width=\columnwidth]{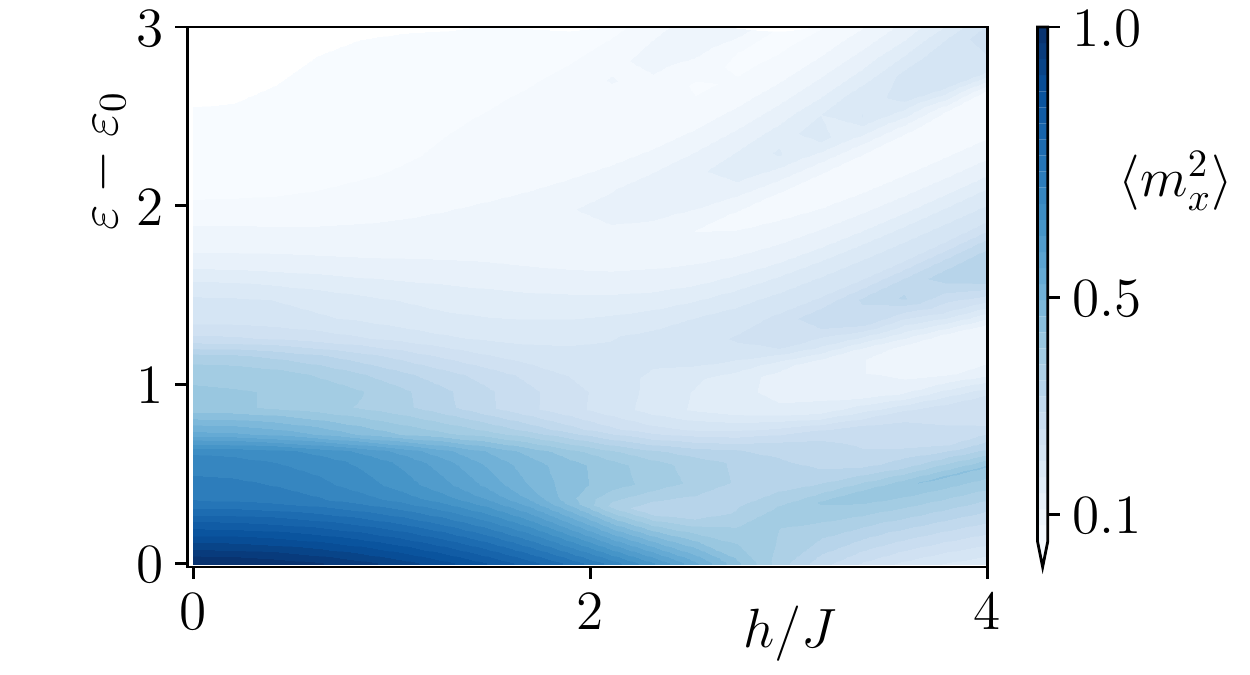}
\caption{Phase diagram of the transverse field Ising model $N=14$ with the interaction coefficients $J_{ij}$ computed according to Eq.  \eqref{eq:J_mn}. All the qualitative features are in good agreement with those shown in Fig. 3 (a) of the main text.}
\label{Fig_realistic} 
\end{figure}

\section{Eigenstate thermalization hypothesis}
\label{sec:eigenst-therm-hypoth}

As explained in the main text, the ETH asserts a specific form of matrix
elements of few-body observables in the energy eigenbasis of an ergodic
many-body Hamiltonian~\cite{Srednicki1999}:
\begin{equation}
  \label{eq:ETH}
  \braket{\ell'|\hat{O}|\ell}=O(\bar{E})\delta_{\ell'\ell}+e^{-S(\bar{E})/2}f_{\hat{O}}(\bar{E},\omega)R_{\ell'\ell}.
\end{equation}
In particular, diagonal and off-diagonal matrix elements are determined by
functions $O(\bar{E})$ and $f_{\hat{O}}(\bar{E},\omega)$, respectively, which
depend smoothly on their arguments $\bar{E}=(E_{\ell}+E_{\ell'})/2$ and
$\omega=E_{\ell'}-E_{\ell}$. $S(\bar{E})$ is the thermodynamic entropy at the
mean energy $\bar{E}$, and $R_{\ell'\ell}$ is a random number with zero mean and
unit variance. From the above form of matrix elements it follows that single
energy eigenstates $\ket{\ell}$ encode thermodynamic properties such as phases
and phase transitions which we typically associate with a microcanonical or
canonical ensemble describing systems in thermodynamic equilibrium. In the main
text, we describe how qualitative aspects of eigenstate thermalization can be
probed across the ferromagnetic phase transition of the long-range transverse
Ising model at finite energy density. Here, we elaborate on direct and more
quantitative experimental tests for diagonal and off-diagonal matrix elements.

\subsection{Diagonal matrix elements}
\label{sec:diag-matr-elem}

\begin{figure}[t!]
\includegraphics[width=0.9\columnwidth]{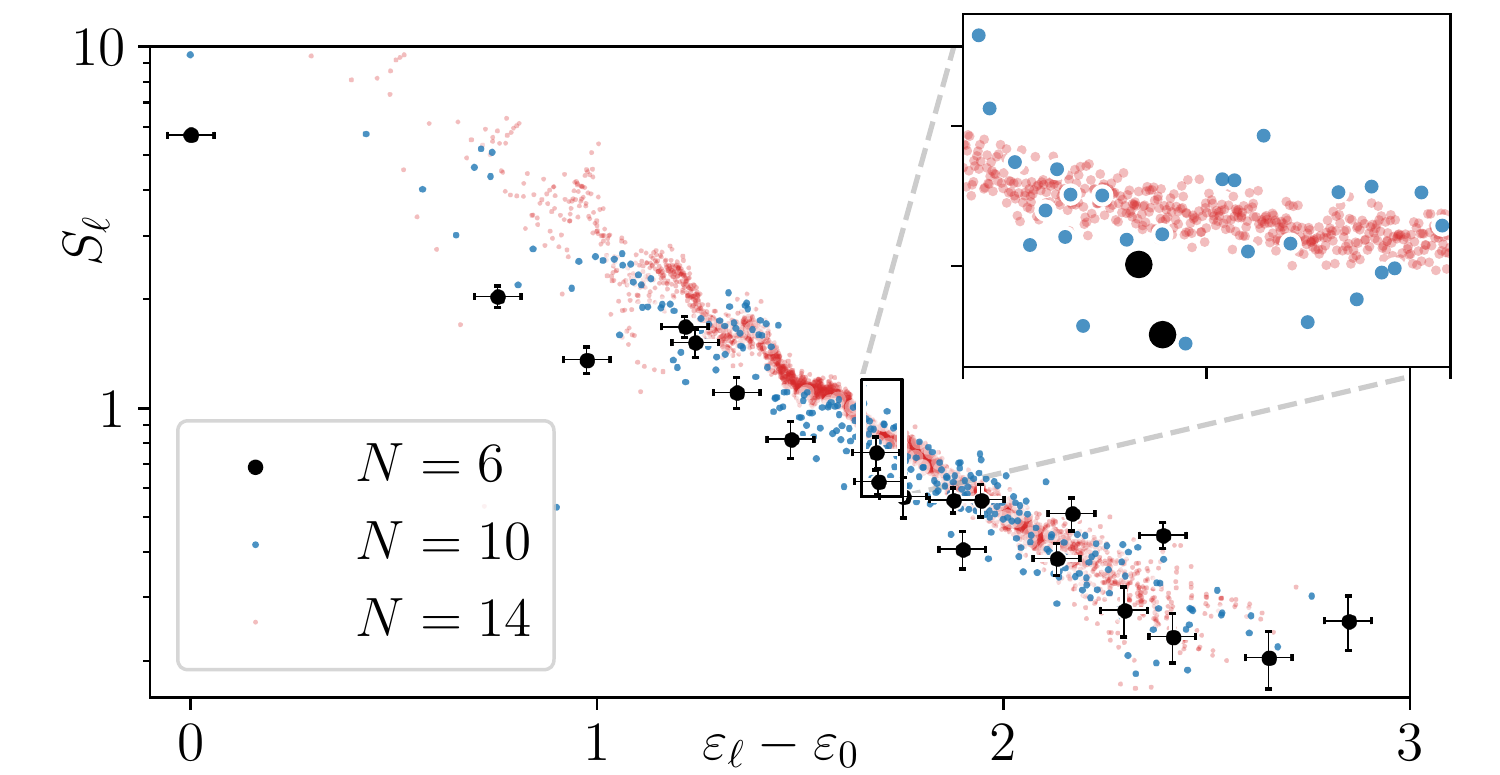}

\caption{Structure factor for single energy eigenstates for $N=6,10,14$ (black,
blue, red dots). The narrowing of fluctuations of the eigenstate expectation
values with increasing system size is a clear indication of the occurrence
of eigenstate thermalization. The Ising model parameters are $\alpha=1.5$
and $h/J=0.75$. The horizontal error bars indicate the estimated energy 
resolution for 6 spins. The vertical error bars provide an estimate of the result
 dispersion for 100 simulated experimental runs per eigenvalue.}
\label{FIG_eth}
\end{figure}

Assessing ETH quantitatively requires to show that fluctuations of
single-eigenstate expectation values $\braket{\ell|\hat{O}|\ell}$ around the
microcanonical average $\tr(\hat{O}\hat{\rho}_{E_{\ell}}^{\mathrm{mc}})$ are
suppressed with increasing system size~\cite{DAlessio2016}. Suitable expectation
values for this purpose are fluctuations of the magnetization,
$\langle\ell|\hat{m}_{x}^{2}|\ell\rangle$ where
$\hat{m}_{x}=N^{-1}\sum_{j}\hat{\sigma}_{j}^{x}$, and the structure factor,
$S_{\ell}\equiv N\langle\ell|\hat{m}_{x}^{2}|\ell\rangle$, which remain finite
in the thermodynamic limit in the ordered and disordered phase,
respectively. Using these quantities, numerical tests of ETH have been performed
for the two-dimensional transverse Ising model with nearest-neighbor
interactions~\cite{Mondaini2016}, and in the one-dimensional
model~\cite{Fratus2016}. In experiments with the trapped ion toolbox, the system
sizes for which single eigenstates can be prepared are limited by the increasing
measurement time which is required to resolve many-body energy level splittings.
Since for intermediate system sizes the number of states in the disordered phase
exceeds the one in the ordered phase (see Fig.~3 in the main text), the most
promising prospect to test ETH quantitatively in experiments is to consider the
structure factor $S_{\ell}$ in the disordered phase. We demonstrate the
narrowing with system size of state-to-state fluctuations of the structure
factor for the one-dimensional transverse Ising model numerically in
Supplementary Fig.~\ref{FIG_eth}. Even for $6$ spins, for which we expect experiments with
current technology to be able to resolve individual energy eigenstates as
discussed in Supplementary Note~\ref{sec:numbers} above, the structure factor exhibits
relatively small state-to-state fluctuations, consistent with ETH. In the
figure, we also indicate the estimated energy resolution for $6$ spins, and the
expected error in determining the structure factor from 1000 measurements. A
further suppression of state-to-state fluctuations is clearly visible for 10 and
14 spin systems, which, however, require improved experimental coherence times
and detection efficiencies. 


\subsection{Off-diagonal matrix elements}
\label{sec:off-diag-matr-elem}

\begin{figure}
\includegraphics[width=1\columnwidth]{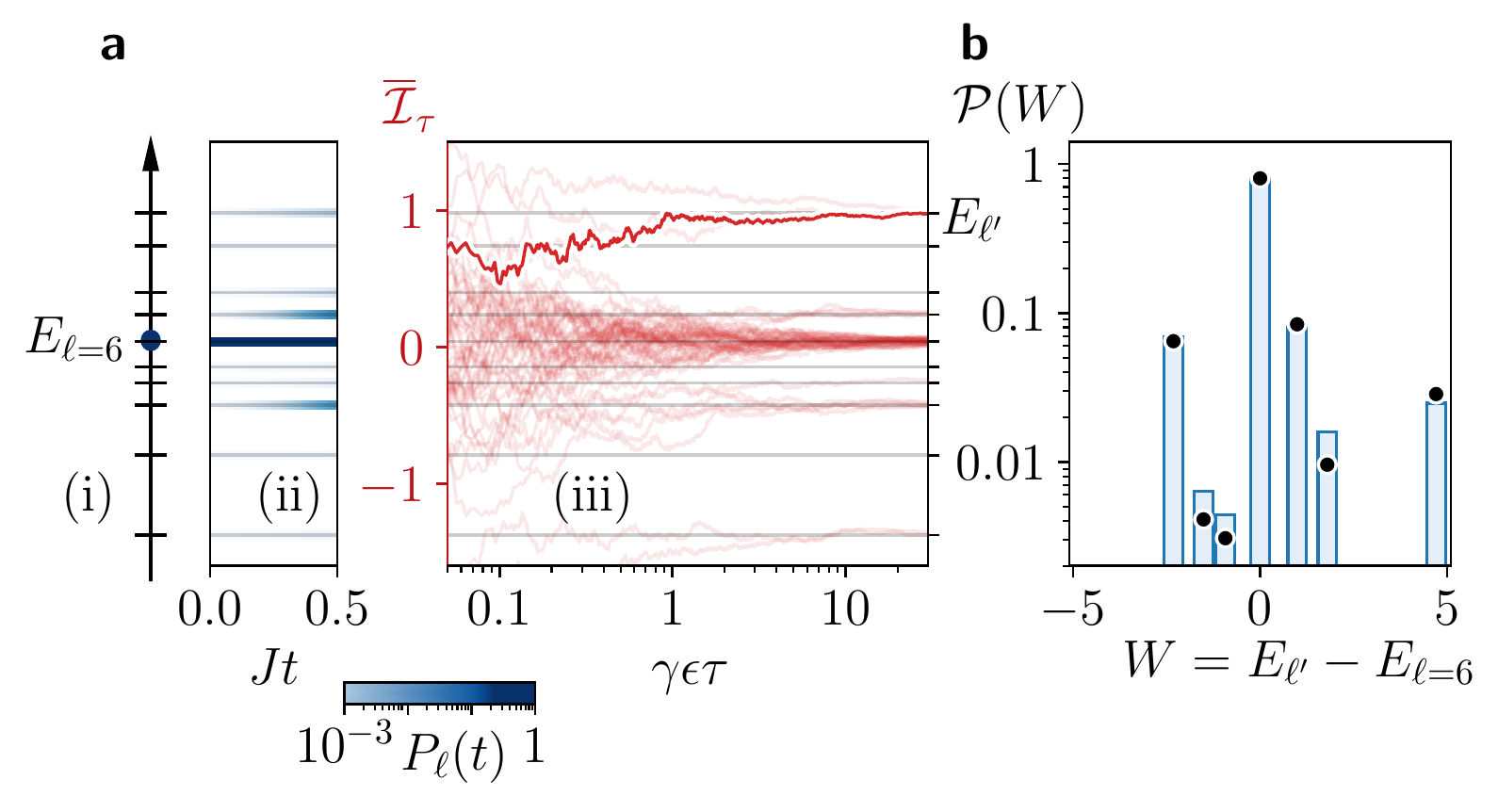} \caption{Measuring off-diagonal
  matrix elements of a local observable
  $\hat{V}=\tilde{h}\,\hat{\sigma}_{j}^{z}$
in the energy eigenbasis via the work probability distribution $P(W)$.
(a) Protocol to measure the work distribution $P(W)$ as described
in the text. (i) The system of $N=5$ spins is prepared in an energy
eigenstate $\ket{\ell=6}$ with energy $E_{\ell=6}$. (ii) By applying
the perturbation $\hat{V}$ at the middle spin ($j=3$ and $\tilde{h}=J$)
during a time $J\Delta t=0.5$, the system is driven into a superposition
of eigenstates $\ket{\ell}$ with probabilities $P_{\ell}(t)$ (blue
shading). (iii) A second measurement of energy collapses the state
of the system continuously to a final state $\ket{\ell'}$. An exemplary
trajectory is shown in dark red. Repeating steps (i), (ii), and (iii)
produces a sample of trajectories (light red), which gives access
to the full distribution $P(W)$ of work $W=E_{\ell'}-E_{\ell}$.
(b) Normalized work distribution $\mathcal{P}(W)=P(W)/\sum_{W}P(W)$
(blue columns). For weak perturbations, the work distribution is determined
by off-diagonal matrix elements $\braket{\ell'|\hat{V}|\ell}$ of
the perturbation. The corresponding approximation to $\mathcal{P}(W)$
is indicated by black dots.}
\label{fig:13}
\end{figure}

Off-diagonal matrix elements are encoded in the dynamics, e.g. in
transition probabilities between energy eigenstates in response to
a weak perturbation. Using the trapped-ion QND toolbox, these transition
probabilities are accessible through the protocol, which is illustrated
in Supplementary Fig.~\ref{fig:13}(a): The preparation (i) of an eigenstate of
$\hat{H}$ with energy $E_{\ell}$ at a given value $h$ of the transverse
field is followed by a period (ii) of length $\Delta t$ of free evolution
{[}$\vartheta=0$ in Eq.~\eqref{eq:H_QND}{]} according to a perturbed
Hamiltonian $\hat{H}'=\hat{H}+\hat{V}$; this is followed by another
measurement (iii) of $\hat{H}$ which yields a value $E_{\ell'}$.
The measurement outcomes determine the work $W=E_{\ell'}-E_{\ell}$
performed on the system by the perturbation $\hat{V}$. To the lowest
order in the perturbation, the work distribution is determined by
off-diagonal matrix elements $\braket{\ell'|\hat{V}|\ell}$ in the
energy eigenbasis, $P(W)=\delta_{\ell'\ell}+\left(\Delta t\right)^{2}\lvert\braket{\ell'|\hat{V}|\ell}\rvert^{2}$.
For $\lvert\braket{\ell'|\hat{V}|\ell}\rvert,W\ll(\Delta t)^{-1}$,
we find good agreement between the exact work distribution and the
lowest-order approximation as illustrated in Supplementary Fig.~\ref{fig:13}(b).
%
%




\bibliographystyle{naturemag}

\end{document}